\documentclass[conference]{IEEEtran}
\IEEEoverridecommandlockouts

\usepackage{cite}
\usepackage{amsmath,amssymb,amsfonts}
\usepackage{algorithm}
\usepackage[indLines=false,commentColor=teal]{algpseudocodex}
\usepackage{graphicx}
\usepackage{textcomp}
\usepackage{xcolor}
\usepackage[hyphens]{url}
\usepackage{tikz}
\usetikzlibrary{quantikz2}
\usetikzlibrary{svg.path}
\usepackage{braket}
\usepackage{subcaption}
\usepackage{booktabs}
\usepackage{comment}
\usepackage{scalerel}

\definecolor{orcidlogocol}{HTML}{A6CE39}
\tikzset{
  orcidlogo/.pic={
    \fill[orcidlogocol]
    svg{M256,128c0,70.7-57.3,128-128,128C57.3,256,0,198.7,0,128C0,57.3,57.3,0,128,0C198.7,0,256,57.3,256,128z};
    \fill[white] svg{M86.3,186.2H70.9V79.1h15.4v48.4V186.2z}
    svg{M108.9,79.1h41.6c39.6,0,57,28.3,57,53.6c0,27.5-21.5,53.6-56.8,53.6h-41.8V79.1z
    M124.3,172.4h24.5c34.9,0,42.9-26.5,42.9-39.7c0-21.5-13.7-39.7-43.7-39.7h-23.7V172.4z}
    svg{M88.7,56.8c0,5.5-4.5,10.1-10.1,10.1c-5.6,0-10.1-4.6-10.1-10.1c0-5.6,4.5-10.1,10.1-10.1C84.2,46.7,88.7,51.3,88.7,56.8z};
  }
}

\newcommand\orcidicon[1]{\textsuperscript{\href{https://orcid.org/#1}{\mbox{\scalerel*{
          \begin{tikzpicture}[yscale=-1,transform shape]
            \pic{orcidlogo};
          \end{tikzpicture}
}{|}}}}}

\usepackage[hyperfootnotes=false,hidelinks]{hyperref}

\def\BibTeX{{\rm B\kern-.05em{\sc i\kern-.025em b}\kern-.08em
T\kern-.1667em\lower.7ex\hbox{E}\kern-.125emX}}
\begin{document}

\pdfpagewidth=8.5in
\pdfpageheight=11in

\newcommand{\revision}[1]{{#1}}
\newcommand{\revisionbox}[1]{{#1}}

\newcommand*\circled[1]{\raisebox{.5pt}{\textcircled{\raisebox{-.9pt} {#1}}}}

\makeatletter %
\newcommand{\linebreakand}{%
\end{@IEEEauthorhalign}
\hfill\mbox{}\par
\mbox{}\hfill
\begin{@IEEEauthorhalign}
}
\makeatother %

\newcommand\blfootnote[1]{%
  \begingroup
  \renewcommand\thefootnote{}\footnote{#1}%
  \addtocounter{footnote}{-1}%
  \endgroup
}

\pagenumbering{arabic}

\title{Optimizing Logical Mappings for Quantum Low-Density Parity Check Codes}
\author{
  \IEEEauthorblockN{Sayam
  Sethi\IEEEauthorrefmark{1}\IEEEauthorrefmark{2}\orcidicon{0009-0005-3056-5285}}
  \IEEEauthorblockA{Electrical and Computer Engineering\\The
  University of Texas at Austin}
  \and
  \IEEEauthorblockN{Sahil
  Khan\IEEEauthorrefmark{1}\orcidicon{0009-0000-4160-8010}}
  \IEEEauthorblockA{Electrical and Computer Engineering\\Duke University}
  \and
  \IEEEauthorblockN{Maxwell Poster\orcidicon{0009-0006-0144-6305}}
  \IEEEauthorblockA{Electrical and Computer Engineering\\The
  University of Texas at Austin}
  \and
  \linebreakand{}
  \IEEEauthorblockN{Abhinav Anand\orcidicon{0000-0002-8081-2310}}
  \IEEEauthorblockA{Electrical and Computer Engineering\\Duke University}
  \and
  \IEEEauthorblockN{Jonathan Mark Baker\orcidicon{0000-0002-0775-8274}}
  \IEEEauthorblockA{Electrical and Computer Engineering\\The
  University of Texas at Austin}
  \linebreakand{}
}

\maketitle
\thispagestyle{plain}
\pagestyle{plain}

\begin{abstract}

  Early demonstrations of fault tolerant quantum systems have paved
  the way for logical-level compilation.
  For fault-tolerant applications to succeed, execution must finish
  with a low total \revision{program} error rate \revision{(i.e., a
  low program failure rate) relative to the baselines}.
  In this work, we study a promising candidate for future
  fault-tolerant architectures with low spatial overhead: the Gross
  code (part of the Bivariate Bicycle code family\cite{tourdegross, bbcodes}).
  Compilation for the Gross code entails compiling to Pauli Based
  Computation and then reducing the pauli-based rotations and
  measurements to Bicycle instructions, including idling operations,
  in-module + inter-module measurements, and $T$-state preparation/injection.
  Depending on the configuration of modules and the placement of code
  modules on hardware, one can reduce the amount of resulting Bicycle
  instructions and improve compilation to produce a lower overall error rate.

  We find that NISQ-based, \revision{and existing FTQC} mappers are
  insufficient for mapping logical qubits on Gross code architectures
  because \circled{1} they \revision{do not account for} the
  two-level nature of
  the logical qubit mapping problem, which separates into code modules
  with distinct measurements, and \circled{2} they naively account only for
  length two interactions, whereas Pauli-Product rotations are up to
  length $n$, where $n$ is the number of logical qubits in the circuit.
  For these reasons, we introduce a two-stage pipeline
  that first uses hypergraph partitioning to create in-module mappings
  (clusters), and then executes a priority-based algorithm to
  efficiently assign clusters onto hardware.
  We find that \revision{our mapping policy reduces} the error contribution from
  inter-module measurements, the largest source of error in the Gross
  Code, by up to $\sim 36\%$ in the best case, with an average
  reduction of $\sim 13\%$ \revision{relative to baselines}.
  \revision{On average, we reduce the failure rates from
    inter-module measurements by $\sim 22\%$ with localized factory
  availability, and by $\sim 17\%$ on grid architectures}, allowing
  hardware developers to be less constrained in developing scalable
  fault tolerant systems due to software driven \revision{reductions in
  program failure rates}.

\end{abstract}

\blfootnote{\IEEEauthorrefmark{1}denotes equal contribution}
\blfootnote{\IEEEauthorrefmark{2}\href{mailto:sayams@utexas.edu}{sayams@utexas.edu}}

\section{Introduction}

Quantum computers have wide-ranging applications across a number of
scientifically interesting problems, including factoring,
optimization, and scientific simulations~\cite{shor1999polynomial,
  kivlichan2020qsimelectrons, childs2018firstqsim, harrow2009hhl,
reiher2017nitrogenfixation}.
The qubits used to perform computation are highly susceptible to
errors; therefore, quantum error correction (QEC) must be implemented
to ensure successful program execution.
QEC codes use many noisy physical qubits to encode information into
reliable \textit{logical} qubits, enabling the construction of
fault-tolerant architectures.

Current quantum computers are rapidly approaching the early fault
tolerant era \cite{googleqecdemo, atomcomputingdemo, QuErademo,
tourdegross}, with demonstrations showing that noisy physical qubits
can be used to achieve lower logical error rates.
This transition is marked by significant progress across different
hardware platforms, with proposals for systems featuring
architectures with thousands of
qubits~\cite{Castelvecchi2023IBM,Pause_2024}, and neutral atom
platforms demonstrating support for up to 6000 physical
qubits~\cite{6100qubits}.
In this regime, the viability of a fault tolerant application is
determined by its overall \revision{program error} rate, and
operational overhead is determined by the cost of performing
\textit{fault tolerant} primitives.
These include syndrome extraction (performed between all operations
to ensure that the physical qubits remain in the codespace), Clifford
logical operations (which safely manipulate the \textit{logical}
state of the code with bounded error overhead), and on-Clifford
T-state injections.
Together, these operations form a universal fault tolerant gate set,
and each QEC code provides its own methods for implementing these
logical operations.

\begin{figure*}
  \centering
  \begin{subfigure}[b]{0.24\textwidth}
    \centering
    \includegraphics[width=0.75\linewidth]{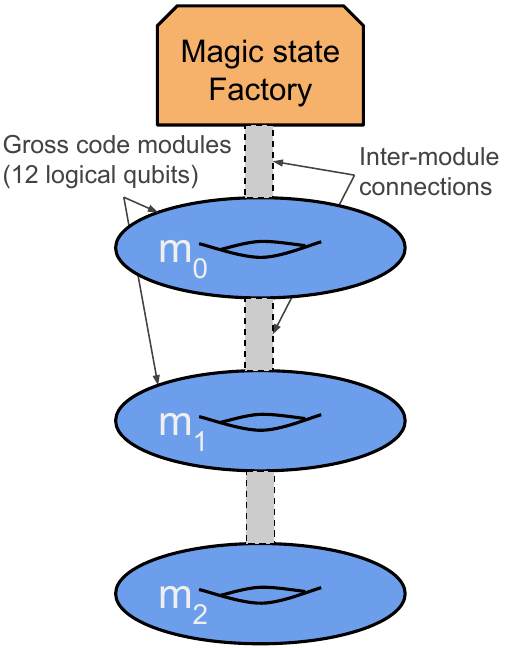}
    \caption{The gross code architecture.}\label{fig:intro-fig-a}
  \end{subfigure}%
  \hfill
  \begin{subfigure}[b]{0.74\textwidth}
    \centering
    \includegraphics[width=\linewidth]{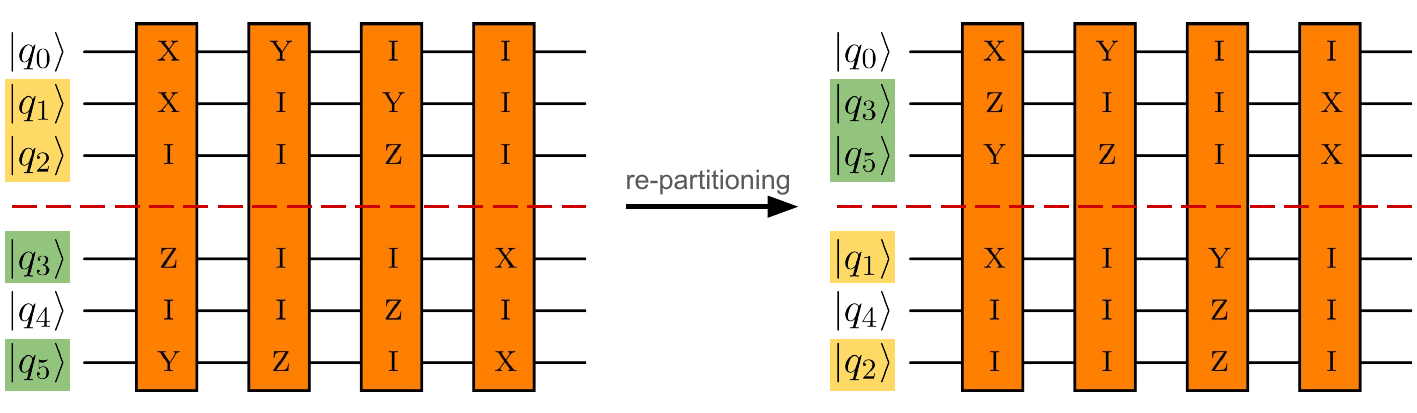}
    \begin{minipage}{.1cm}\vfill
    \end{minipage}
    \caption{Re-partition of qubits into different modules, reducing
    number of inter-module measurements.}
  \end{subfigure}%
  \hfill
  \begin{subfigure}[b]{\textwidth}
    \centering
    \revisionbox{\includegraphics[width=\linewidth]{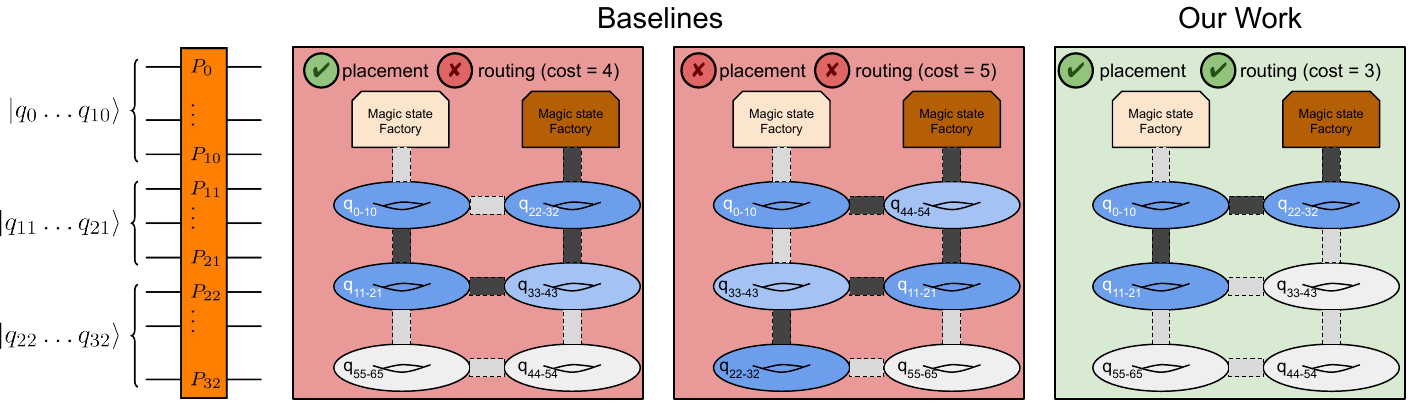}}
    \caption{Our Mapper uses hypergraph partitioning, and a
      priority-based algorithm to assign module
    placements.}\label{fig:intro-fig-c}
  \end{subfigure}
  \caption{
    \textbf{(a)} Three gross-code modules (12 qubits per module)
    connected linearly. Magic states are routed via inter-module
    connections to perform PRRs both within and across gross-code
    modules. \textbf{(b)} Qubits are initially divided into modules
    $\{q_0,q_1,q_2\}$ and $\{q_3,q_4,q_5\}$, requiring 4 inter-module
    interactions to complete the program. After re-partitioning into
    $\{q_0,q_3,q_5\}$ and $\{q_1,q_4,q_2\}$, the number of
    inter-module interactions reduces to only 1. \textbf{(c)} Our
    mapping policy compared to current state-of-the-art baselines.
    Dark blue modules contain qubits associated with the operation,
    while light blue modules are used only for routing to other
    modules. Either baselines perform adequate placement, but neglect
    routing, or vice versa, incurring substantial overhead by using
    more inter-module connections than required. Our mapping policy
    performs improved placement and routing to reduce the number of
    inter-module connections required to perform PPRs, significantly
    improving program success rate and reducing error accumulation.
  }\label{fig:intro-fig}
\end{figure*}

Recent proposals~\cite{constantoverheadqecneutralatoms, tourdegross}
have increasingly focused on developing fault tolerant architectures
around spatially efficient codes, which can encode more logical
qubits per physical qubit, achieving a higher \textit{encoding rate}.
In particular, Bivariate Bicycle (BB) Codes~\cite{bbcodes},
especially the Gross code~\cite{tourdegross}, have been of interest
due to their ability to achieve high distance (which determines the
number of simultaneously correctable errors) while maintaining high
encoding rate.
The BB codes come with a fault tolerant architectural instruction
set, known as the Bicycle Instructions, and several theoretical
architectures based on the Gross code using the bicycle instruction
set have significantly lower resource requirements compared to other
early fault-tolerant proposals~\cite{tourdegross,
he2025extractorsqldpcarchitecturesefficient}. %
These architectures assume that a given quantum circuit is first
compiled according to the \textit{Pauli Based Computing} (PBC) model,
resulting in a list of Pauli Product Rotations (PPR) and Pauli
Product Measurements (PPM).
These are then translated into Bicycle Instructions, which include
\textit{idling operations} (syndrome extraction), \textit{in-module}
and \textit{inter-module} measurements to address qubits within and
across modules, respectively, and $T$-state injections.
For an $n$-qubit Pauli product rotation, inter-module measurements
are generated for every module on which the Pauli string acts
trivially, while in-module measurements performed on each module
handle the 11-qubit Pauli fragments corresponding to each block.
Meanwhile, a T-factory runs \revision{separately at the edge of the
line} to generate T-states for Pauli product rotation of the form
$Rz(\theta)$, and distributes them to the appropriate modules using
inter module measurements from the cultivation and/or factory. %

\textbf{The Logical Qubit Mapping Problem} Gross code blocks
\revision{give the ability to} configure both a module’s internal
qubit layout and its \revision{relative} placement on the device
(assumed to be either a line or \revision{grid} due to hardware
constraints~\cite{IBM2025LargeScaleFTQC, tourdegross}).
Changing this mapping directly affects the number of inter- and
in-module measurements, as well as the distances between modules and
the magic state factory, as seen in Figure~\ref{fig:intro-fig-c}.
Since each instruction contributes to the total \revision{error (or
failure) rate of the program}, the feasibility of running early
fault-tolerant applications hinges on minimizing this total error budget.
Our goal is therefore to generate mappings that reduce the
\revision{total error contribution} of these instructions.

\textbf{Prior Work and Their Limitations} Although this problem
resembles the task of mapping physical qubits onto hardware in the
noisy intermediate scale quantum (NISQ)
era~\cite{zou2024lightsabrelightweightenhancedsabre,
10.1145/3387902.3392617,meqc,max-sat}, \revision{as well as mapping
in FTQC~\cite{autobraid,silva_multi-qubit_2024}}, existing solutions
are inadequate for Gross-code logical mapping for the following reasons:
\begin{enumerate}
  \item \revision{NISQ mappers} model interactions only as length-2
    operations, whereas PPR can involve up to $n$-logical qubits.
  \item NISQ and FTQC mappers don't capture the two-level structure
    of the mapping problem. Namely, the distinction between
    \textit{clustering} arbitrarily long interactions without the
    ability to route efficiently, and \textit{assigning} modules to a
    linear topology in such a way where paths can be reused.
  \item \revision{FTQC mappers assume that the cost of operations
      does not scale with the distance between the associated qubits,
      only requiring a routing path. Yet gross code operations are
      largely blocking, and their error costs scale with the distance
    between qubits/modules.}
\end{enumerate}
To overcome these limitations, we propose a first of its kind logical
qubit mapper tailored for gross code architectures.
Our mapping policy uses a two-step solution to address the in-module
clustering problem and hardware assignment problem separately.
First, our policy uses a hypergraph representation of length-$n$
interactions and performs hypergraph partitioning to construct
improved in-module mappings.
This reduces the total number of inter-module measurements by up to
$\sim 62\%$ in the best case, and by $20\%$ on average.
Next, the resulting clusters are mapped to hardware using a
priority-based heuristic that considers certain activation patterns
of inter-module measurements.
This assignment is based on a logical-level connectivity graph.
\revision{The approach used by our mapping policy is easily
  extensible to other BB codes, as well as other high-rate codes (i.e.
codes with $k > 1$).}

Furthermore, we \revision{perform sensitivity studies to compare} our
proposal's efficacy on \revision{alternative topologies} to the
initial linear one proposed by \cite{tourdegross} and under
consideration by industry \cite{IBM2025LargeScaleFTQC}.
We also develop an improved path-finding algorithm for creating
efficient inter-module measurement paths for \revision{grid}
topologies. We find that, across the \revision{dynamic} part of
compilation, We achieve \revision{over $40\%$} reduction in program
\revision{failure} error rate in the best case, and a
\revision{$17\%$} reduction on average compared to the main baseline,
SABRE~\cite{zou2024lightsabrelightweightenhancedsabre,SABRE}.
\revision{We also perform sensitivity studies of our proposal with
  variable factory availability and observe up to a $\sim 75\%$
  reduction in program failure rate, with an average reduction of $\sim
23\%$ when each module has an associated factory.}
These are noticeable improvements that can bring hardware designers
closer to fault tolerance without any hardware alterations.

Our contributions are as follows:
\begin{enumerate}
  \item We show that NISQ mappers cannot effectively solve the
    logical qubit mapping problem on BB codes because of the
    two-level nature of the problem and the presence of $n$-qubit interactions.
  We introduce a two-stage algorithm that 1) performs hypergraph
partitioning for up to $n$-qubit interactions and 2) executes a
priority-based heuristic that encourages low-cost, repeated
inter-module measurement patterns for hardware assignment.
\item We evaluate our algorithm on another realistic logical
connection topology, the long-grid, and propose an efficient
path-finding algorithm based on the topology.
\item We demonstrate that our mapping strategy reduces
\revision{program failure rates} by noticeable amounts,
\revision{averaging $\sim 4\%$ at current estimates of error rates
  purely through software improvements, and noting relative reduction
will improve as devices get better}, thereby easing hardware
requirements for achieving scalable early fault tolerance.
\end{enumerate}

\section{Background}

\subsection{Quantum Error Correction}
Quantum computers are inherently noisy and this noise affects the
accuracy, or fidelity, of quantum computations.
Quantum Error Correction (QEC) is a technique that protects program
execution from errors by adding redundancy to the encoded information.
A quantum error correcting code (QECC) is defined by its code
parameters $[[n, k,d]]$, where $n$ is the number of physical qubits
that are used to encode $k$ logical qubits and $d$ is the code distance.
The code distance $d$ is an important metric that determines the
error-correcting capability of the code.
A code of distance $d$ can correct up to $\lfloor (d-1)/2 \rfloor$
errors on distinct physical qubits.
Another important metric of a QEC code is its encoding rate, which is
defined as $k/n$.
For the same distance, a QECC with a higher encoding rate is
preferred over one with a lower rate, as it uses fewer physical
qubits to encode the same number of logical qubits, thereby improving
the space efficiency of the quantum computer.
A prominent code family that satisfies these desirable properties is
the class of quantum low-density parity-check (qLDPC) codes.

\subsubsection{Gross Code}
The gross code is a member of the family of the Bivariate Bicyclic
codes, which is a subset of the qLDPC codes. The gross code has
parameters $[[144, 12, 12]]$.
Therefore, each logical block in the gross code consists of 144
physical qubits that encode 12 logical qubits, with a code distance of 12.
Another variant of the code, the two-gross code, has code parameters
of $[[288, 12, 18]]$.
Both these codes are promising for large-scale fault-tolerant quantum
computation (FTQC) due to their high encoding rates, and performing
logical operations on gross (and two-gross) codes has been widely
studied in literature~\cite{vf7v-cpq9, 10812769}.

\subsection{Measurement-Based/Pauli-Based Computation}
Measurement-based quantum computing (MBQC)\cite{PhysRevLett.86.5188,
PhysRevResearch.2.033305} is a paradigm of quantum computation in
which computation proceeds entirely via projective measurements and
resource-state initialization
Pauli-Based Computation (PBC)\cite{PhysRevX.6.021043} is a
specialized case of MBQC where the projective measurements are along
arbitrary Pauli bases.
To achieve universal quantum computation, PBC additionally requires
access to a non-Clifford resource state which is typically either the
$\ket{T}$ state ($= T\ket{+}$ state), or the $\ket{m_{\theta}}$ state
($= Rz(\theta)\ket{+}$ state).
This paradigm is particularly attractive for architectures that do
not natively support a universal gate set, or for QECC, such as the
qLDPC code, for which a universal logical gate set is not yet
known~\cite{tourdegross}.
However, PBC introduces the challenge of reliably synthesizing magic
states with high fidelities, to ensure that the overall quantum
circuit executes with sufficiently high fidelities.

\subsubsection{Translating to PBC Circuits}\label{sec:pbc-translation}
\begin{figure}
  \centering
  \includegraphics[width=\linewidth]{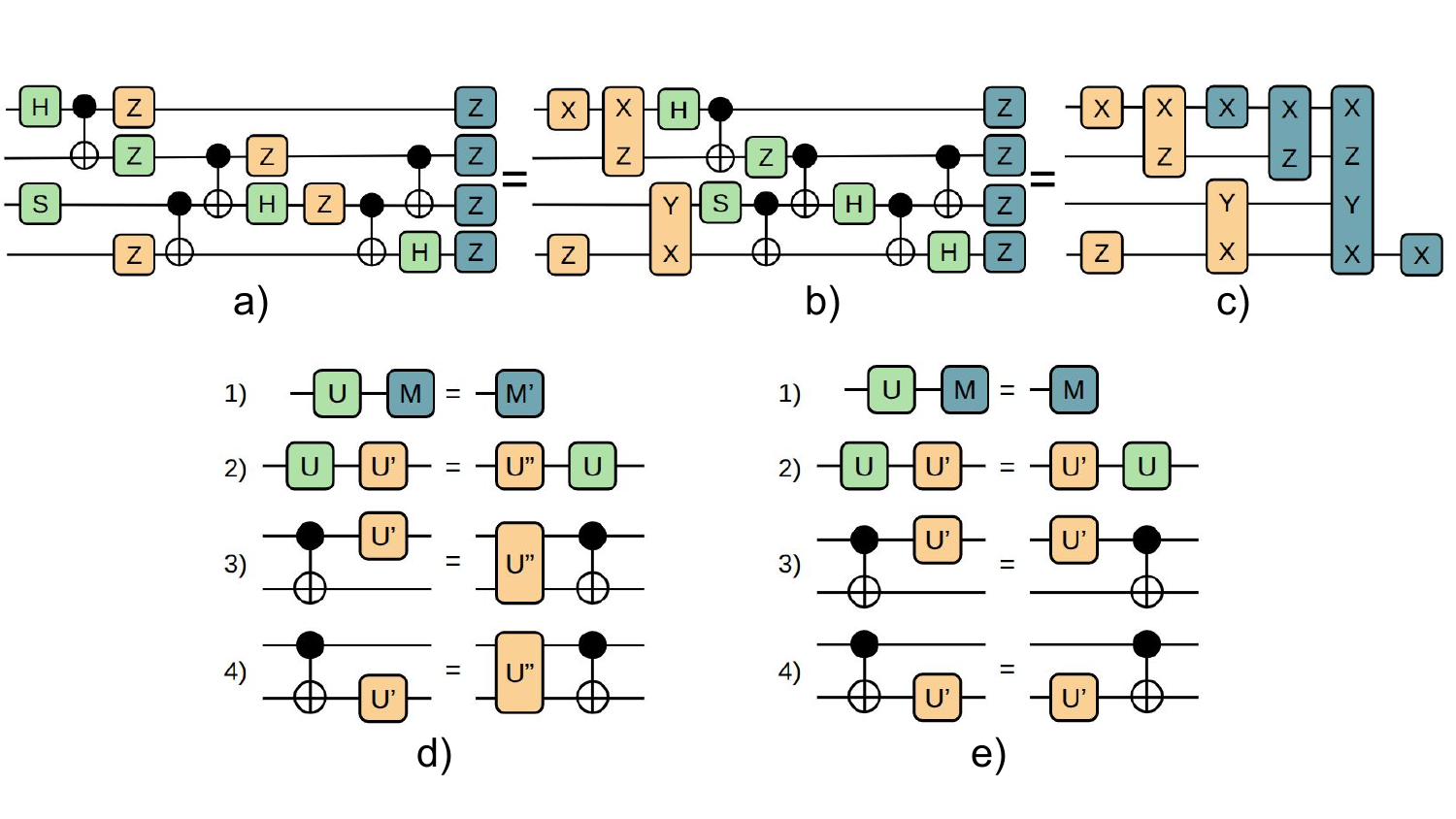}
  \caption{A schematic illustrating the compilation of a circuit into
    its Pauli-Based Computation (PBC) form. (a) A circuit transpiled
    into H, S, CX, and T (or Rz) gates. Non-Clifford gates are shown in
    orange, Clifford gates in green, and measurements in blue. The
    anticommutation rule (d) and commutation rule (e) (see
    Ref.~\cite{litinski2019gameofsurfacecodes} for details) are applied
    to systematically commute non-Clifford gates to the front,
    producing the intermediate representation in (b). Applying these
    rules to the measurements results in the fully translated PBC
  circuit in (c).}\label{fig:PBCmeasurement}
\end{figure}
Recent works~\cite{litinski2019gameofsurfacecodes, tourdegross} have
proposed various compilation strategies for efficient translation
from a gate-based circuit to a PBC circuit.
These approaches start with a Clifford + T (or Rz) circuit.
The Clifford gates in the circuit are first converted into Pauli
measurements, which are then commuted across the non-Cliffords,
modifying them as they commute through.
At the end of the process, we are left with a sequence of Pauli
product measurements.
Each of these measurements requires a \textit{magic state}, which is
either a $\ket{T}$ or a $\ket{m_\theta}$ state.
We demonstrate execution of the translation strategy, along with the
commutation rules in Figure~\ref{fig:PBCmeasurement}.

\begin{figure}
  \centering
  \includegraphics[width=0.7\linewidth]{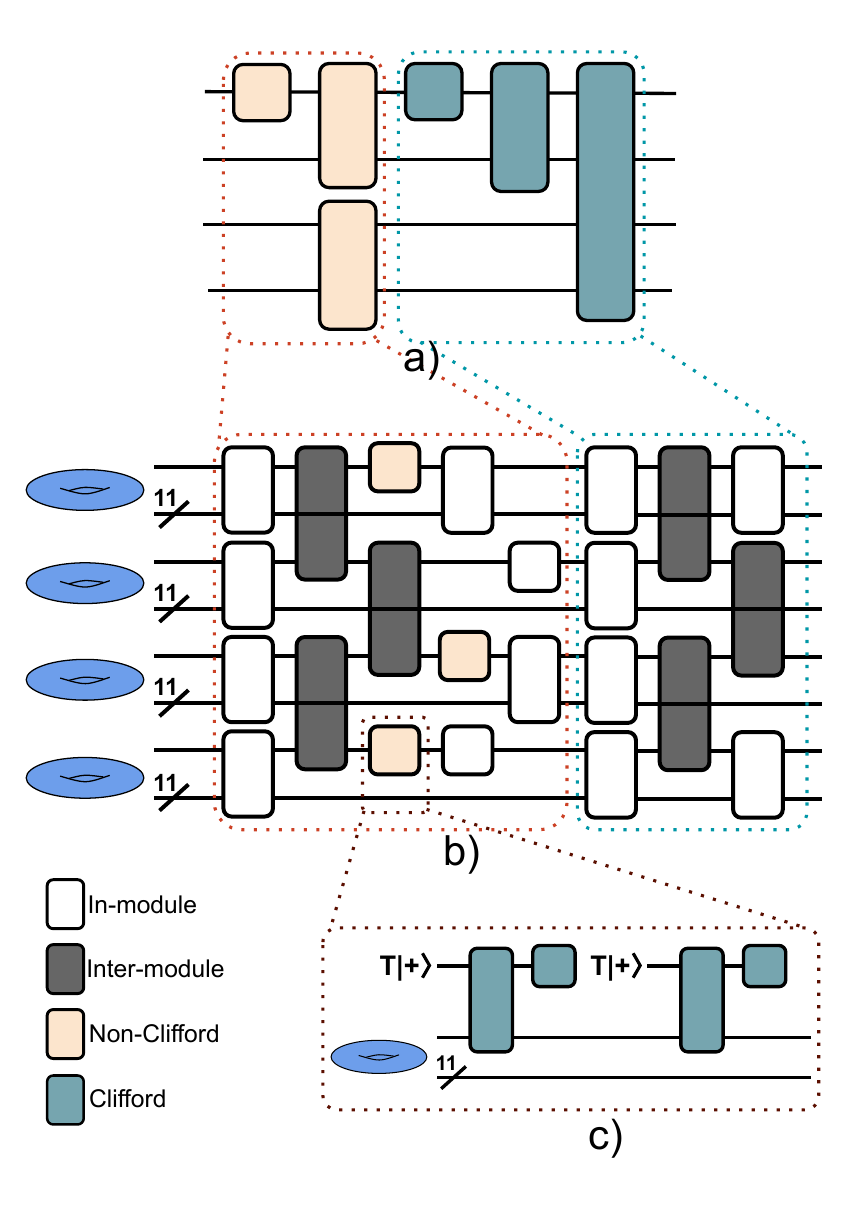}
  \caption{Translation from a Pauli-Based Computation (PBC) circuit
    to the Gross-code Bicycle instruction set. (a) PBC circuit with
    Clifford measurements represented by blue boxes and non-Clifford
    measurements by orange boxes. (b) Resulting instruction sequence
    for execution on the Gross code, including in-module measurements
    (white boxes), inter-module measurements (grey boxes), and
    non-Clifford operations (orange boxes), the latter realized through
    T-state injection shown in (c). See Ref.~\cite{tourdegross} for
  further architectural details.}\label{fig:TDGarchitecture}
\end{figure}

\subsection{Tour de Gross Architecture}\label{sec:tour-de-gross}
Tour de gross~\cite{tourdegross} is a compilation scheme proposed
recently for executing universal, fault-tolerant operations for the gross code.
Their proposed architecture consists of a single magic state factory
at the top, followed by a line of gross code modules.
Each module consists of $12$ logical qubits, where the first logical
qubit is used for routing and is called an ancilla (or a pivot, $p_i$) qubit.
The other $11$ qubits are used to store program qubits and are called
logical program qubits.
The tour de gross architecture supports $4$ different kinds of
operations, namely the shift automorphisms, in-module measurements,
inter-module measurements and T injections.
We list their properties in Table~\ref{tab:error-rates}.
An illustration of translation of a circuit in the PBC model to the
Bicycle instruction set is shown in Figure~\ref{fig:TDGarchitecture}.
Note that Compiling to $Rz(\theta)$ rotations in the Bicycle
architecture results in PPRs of different widths than compiling directly to $T$.

\begin{table}
  \setlength\tabcolsep{0pt}
  \centering
  \begin{tabular*}{\columnwidth}{@{\extracolsep{\fill}}lcccc}
    \toprule
    Instruction & Notation & Timesteps & \multicolumn{2}{c}{Logical
    error rate ($= P_i$)} \\
    \cline{4-5}
    & $(i)$ & & $p = 10^{-3}$& $p = 10^{-4}$\\
    \midrule
    idle & $I$& $8$ & $10^{-8.8\pm 0.2}$ & $10^{-14.8\pm 0.4}$\\
    shift automorphism & $U$ & $14$ & $10^{-6.4 \pm 0.2}$ &
    $10^{-12.2\pm 0.5}$\\
    in-module meas. & $B$ & $120$ & $10^{-5.0\pm 0.1}$ & $10^{-9.0\pm 0.2}$\\
    inter-module meas. & $C$ & $120$ & $10^{-2.7\pm 0.1}$ & $10^{-7.4\pm 0.3}$\\
    T injection & $T$ & $471$ & $10^{-5.5} + P_C$ & $10^{-7.4} + P_C$\\
    \bottomrule
  \end{tabular*}
  \caption{Bicycle instruction properties: Logical error rates and
    timestep durations at physical error rates $p = 10^{-3}$ and $p =
    10^{-4}$. $P_C$ is the logical error rate of inter-module
  measurements.}\label{tab:error-rates}
\end{table}

\subsubsection{Quantum Program Execution}
Program execution on the tour de gross architecture proceeds as
follows (also shown in Figure~\ref{fig:TDGarchitecture}).
Consider an architecture with $M$ modules laid out in a line, and a
Pauli-based measurement $P(\varphi) = (P_{m_1}\otimes  P_{m_2}
\otimes \cdots P_{m_M})(\varphi)$.
Note that some of these $M$ modules, say $m_i$, may not be involved
in the measurement, and thus will have $P_i = \text{I}^{\otimes 11}$.
To execute $P(\varphi)$, a GHZ state is prepared via inter-module
measurements on the pivot qubit.
Within each module $m_i$, the entangled pivots then interact with the
logical program qubits via in-module measurements to project the
qubits along the $P_{m_i}$ bases.
Finally, the $\ket{m_\varphi}$ state prepared using the magic state
factory and injections onto the pivot present in $m_1$, the module
closest to the factory.

\begin{figure*}
  \centering
  \revisionbox{\includegraphics[width=\linewidth]{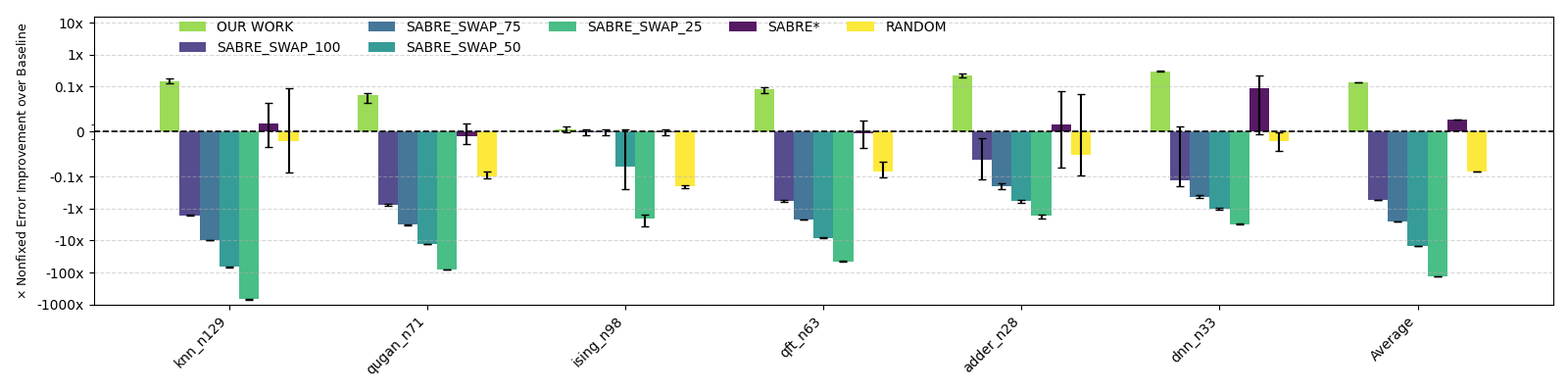}}
  \caption{\revision{A plot showing how SWAP insertion leads to a
      substantial increase in program error rate, contrary to NISQ
      compilation schemes. We evaluate the SABRESWAP strategy by
      inserting SWAPs every $f$ rotations, and vary $f$ between $25$ and
      $100$ rotations at intervals of $25$ rotations. Smaller $f$ leads
      to a much larger increase in inter-module measurements, which also
      makes it infeasible to simulate SABRESWAP strategies on all
  benchmark suites, or with a higher number of trials.}}\label{fig:sabreswap}
\end{figure*}

\subsection{Qubit Mapping Problem}
The problem of mapping program qubits onto the device has been widely
studied in both the NISQ and FTQC
regimes~\cite{zou2024lightsabrelightweightenhancedsabre,SABRE,10.1145/3387902.3392617,autobraid,meqc}.
The NISQ-based approaches, such as
SABRE\cite{zou2024lightsabrelightweightenhancedsabre,SABRE,10.1145/3387902.3392617,meqc},
solve the mapping problem by determining an initial qubit layout, and
inserting SWAP gates to enable routing.
However, as discussed in Section~\ref{sec:tour-de-gross}, executing
SWAP gates in PBC is expensive, and therefore, NISQ-based mapping
approaches provide suboptimal layouts for execution.
The algorithms used for NISQ-based mapping also scale poorly with an
increase in the number of qubits and gate depths, making them
unsuitable for large-scale FTQC circuit compilation.
Moreover, NISQ-based compilation does not account for the location of
magic-state distillation factories, which are located at the boundary
of the logical device, as there is no notion of factories in
NISQ-based computation.

\par The current state-of-the-art FTQC mapping approach
\cite{autobraid}, on the other hand, supports long-range gates, and
therefore does not need to insert SWAP gates for routing.
They instead attempt to maximize parallel execution of the long-range
gates by partitioning the qubits into groups that can be operated on
simultaneously.
PBC compilation, and in particular, the tour-de-gross architecture,
does not support parallel execution of Pauli-based measurements.
Therefore, FTQC-based mapping approaches that aim to maximize
parallelism do not work well for PBC architectures either.

\par The translation strategies discussed in
Section~\ref{sec:pbc-translation} merely translate the input circuit
into a PBC circuit and do not address the the logical qubit mapping problem.
Figure~\ref{fig:intro-fig-c} shows how poor mappings can lead to
increase in error rates and execution times.
We discuss this in detail in Section~\ref{sec:priority-assignment}.

\section{Identifying Constraints of Logical Qubit Mapping}

In this section, we explore how different measurement types are
affected by a logical qubit mapping.
We first show that neither SWAP insertion nor in-module measurement
substantially improve error rates.
We also demonstrate that the majority of the \revision{program error
(failure)} is dominated by fixed, synthesis-based, inter-module cost,
and conclude that the most effective strategy for reducing
\revision{program error (failure)} minimizing the non-synthesis-based
inter-module measurements.
We explore two methods to reduce inter-module measurements, and later
use these principles to build our mapping policy.

\subsection{Routing via SWAP Insertion is Impractical}\label{sec:swap-bad}
We can execute a Pauli measurement on qubits $q_A, q_B$ in modules
$m_i, m_j$ by first swapping $q_A$ with another logical qubit in
$m_j$, or swapping $q_B$ with another logical qubit in $m_i$. We can
then execute the target Pauli measurement via a sequence of in-module
measurements. In either case, we will need to execute $3$ CNOT gates
that act on $q_A, q_B$. Each CNOT gate requires entangling modules
$m_i, m_j$, and all the modules that are present between the two.
Therefore, if $m_i, m_j$ are $x$ modules apart, this introduces $3x$
inter-module measurements. Additionally, a sequence of in-module
measurements in $m_i$ (or $m_j$) would be required to execute the
Pauli measurement between $q_A, q_B$. Say, this requires $y$
in-module measurements. Since $P_C, P_B$ are small, we can compute
the total error as $\varepsilon_\text{SWAP} \approx 3xP_C + yP_B$. We
require $120\cdot(3x + y)$ timesteps using the SWAP insertion approach.
\par On the other hand, if we would have directly entangled the two
modules $m_i, m_j$ via $x$ inter-module measurements, followed by
in-module measurements within both $m_i$ and $m_j$. This requires
only $x$ inter-module measurements and $y_i + y_j$ in-module
measurements. The error of this method of execution is
$\varepsilon_\text{inter} \approx xP_C + (y_i + y_j)P_B$. This
approach requires $120\cdot(x + \max(y_i, y_j))$ timesteps.
\par Therefore, as long as $\max(y_i, y_j) - y < 2x$, this approach
is faster than the version via SWAP insertion. Since in the
tour-de-gross ISA, over $90\%$ of the in-module sequences are close
to the mean~\cite{tourdegross}, the difference will be very close to
$0$, and the inequality holds. Additionally, since $P_C$ is $2$
orders of magnitude larger than $P_B$ (Table~\ref{tab:error-rates}),
and $y, y_i, y_j \in [1, 25]$, for any $x$ (distance between the
modules), we will always have $\varepsilon_\text{inter} \ll
\varepsilon_\text{SWAP}$. Therefore, routing logical qubits through
SWAP operations is always inferior for the tour-de-gross ISA.
\par \revision{We also estimate the effects of SWAP insertion to
  remap the grid with changing frequencies. In
  Figure~\ref{fig:sabreswap}, we insert SWAPS at varying frequencies
  via the SABRESWAP heuristic. We observe that, regardless of SWAP
  insertion frequency, the program error rates, and equivalently the
total number of inter-module measurements worsen by 1-2 orders of magnitude.}

\subsection{In-Module Measurements are not the Bottleneck}
A mapping fixes blocks of 11 logical qubits within each module for
the logical computation, which entails compiling a list of PPRs $R$
that make up the entire circuit.
A single $n$-qubit rotation $r$ is spread across $M$ modules and the
rotation corresponding to a module $m$ is denoted as $B_{m,r}$.
Therefore, the total in module measurement cost, $C_{in}$,  is

\[C_{in} =  \sum^R_r \sum^M_m T(B_{m,r})\]

where $T$ is the cost associated with the rotation from the lookup
table of optimal Clifford synthesis in Ref.~\cite{tourdegross}.
The cost range from 1-25 in-module measurements, with an average of
18.5 measurements.

\subsubsection{Contrived Case}

In Figure~\ref{fig:in-cost_breakdown}, we show a contrived example
where mapping drastically changes the in-module measurement cost for
a single rotation.
A naive mapping leads to a synthesis of $5$ rotations for the top
module and $7$ rotations for the bottom module, leading to $12$ total
in-module measurements required for the entire rotation.
However, in a mapping that is aware of the costs of Clifford
synthesis, one can re-partition the qubits and perform this rotation
with only $2$ in-module measurements, as both of these rotations are
native measurements to the BB code.

\begin{figure}[htbp!]
  \centering
  \includegraphics[width=\linewidth]{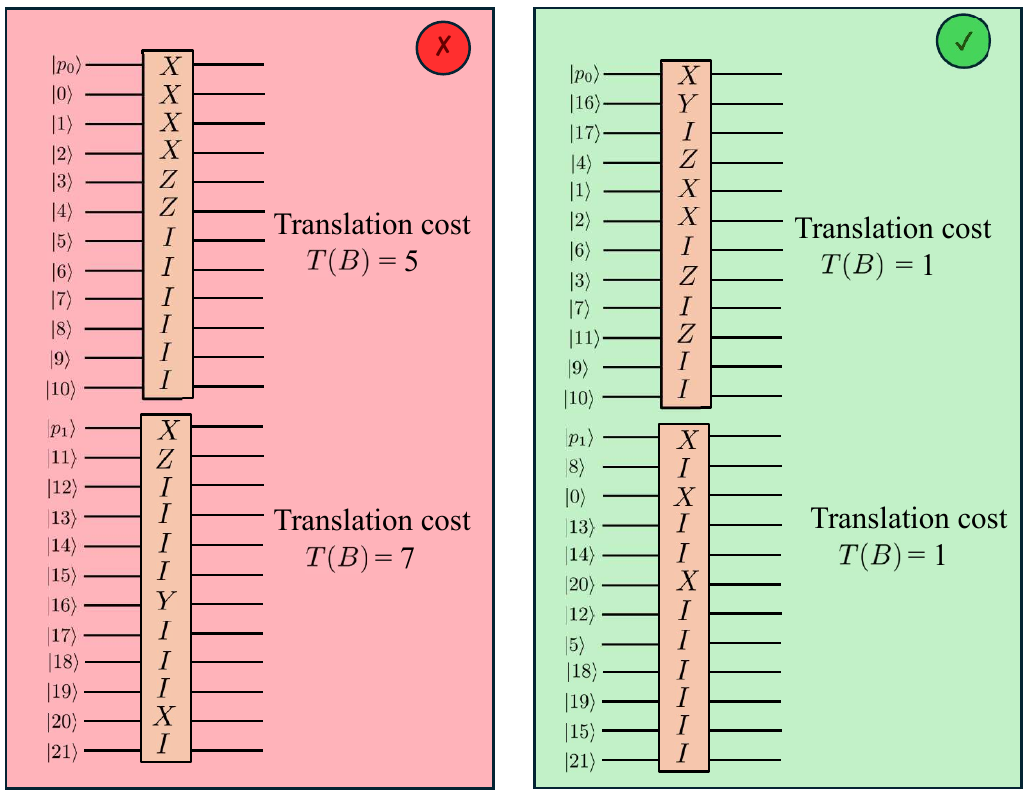}
  \caption{A figure illustrating the effect of logical-qubit
    remapping on in-module measurement cost. Left: A naive mapping
    leads to higher cost. Right: A mapping exploiting the in-module
  measurement Clifford table leads to significantly reduced cost.}
  \label{fig:in-cost_breakdown}
\end{figure}

\subsubsection{Realistic Scenario}
In practice, a circuit is not just one rotation but a list of
rotations, of large depth (up to tens of thousands).
Once a mapping is assigned, it must serve as a \textit{global
solution} across all rotations, since inserting SWAP gates is
sub-optimal as discussed in Section~\ref{sec:swap-bad}.
However, as the number of rotations increases, the in-module
measurement synthesis cost per rotation converges toward the average
of 18.5, and over $90\%$ of the measurements require $19\pm 6$
in-module measurements~\cite{tourdegross}.
Experimentally we observe that the in-module measurement errors lag
behind inter-module errors by about one order of magnitude, as seen
in Figure \ref{fig:cost_breakdown}.
This is because given the error model from
Table~\ref{tab:error-rates} the ratio of in-module to inter-module
error rates is $\approx$ 1:100, and since we have 18.5 in-module
measurements for every inter-module measurement, we see approximately
one order of magnitude lag behind.

\begin{figure}[htbp!]
  \centering
  \includegraphics[width=\linewidth]{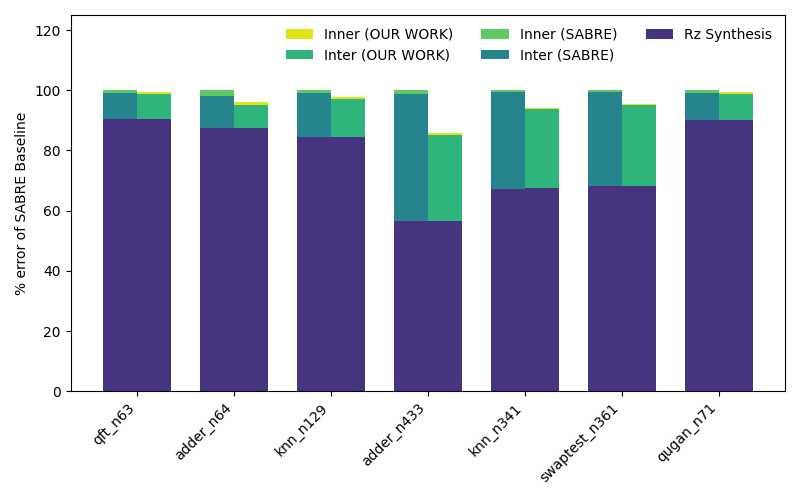}
  \caption{Breakdown of total \revision{program} error contributions
  across different error sources for a range of benchmarks.}
  \label{fig:cost_breakdown}
\end{figure}

Even in an idealized scenario where an in-module measurement could be
traded for an inter-module measurement,
Figure~\ref{fig:inter_expensive} shows that inter-module measurements
remain far more expensive.
For example, the worst-case in-module cost of $25 + 25 = 50$ and one
inter-module measurement still has a lower error rate than $1 + 1 + 1
= 3$ in-module and 2 inter-module measurements.
Therefore, optimizing in-module measurements provides little
advantage, and we shift our focus on reducing the inter-module measurements.

\begin{figure}[htbp!]
  \centering
  \includegraphics[width=\linewidth]{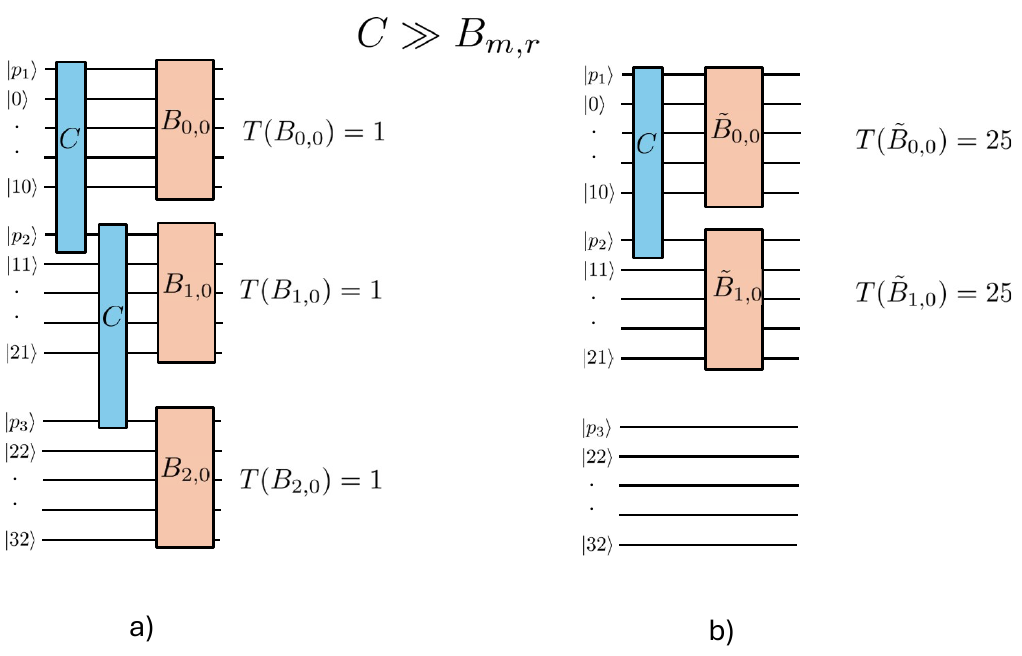}
  \caption{Inter-module measurements (C) are much more expensive than
    in-module measurements ($B_{m,r}$), such that saving just one
    inter-module measurement part (b) is worth the cost of having the
  worst case in translation for in-module measurements (a)}
  \label{fig:inter_expensive}
\end{figure}

\subsection{Inter-Module Measurements can be Minimized via Partitioning}
\label{sec:reduce-intermodule}
Different logical mappings can also vary the number of inter-module
measurements, which are the most expensive bicycle instruction in
terms of error contribution (as seen from Table~\ref{tab:error-rates}
and Figure~\ref{fig:cost_breakdown}).
Notably, we distinguish between the fixed amount of synthesis-based
inter-module measurements, which are needed for each benchmark, and
the non-fixed, or \textit{optimizable}, inter-module measurements
which are mapping/compilation dependent.
We can reduce the number of non-synthesis-based inter-module
measurements through mapping due to two reasons.

\begin{figure*}[ht!]
  \centering
  \begin{subfigure}{\textwidth}
    \centering
    \includegraphics[width=\linewidth]{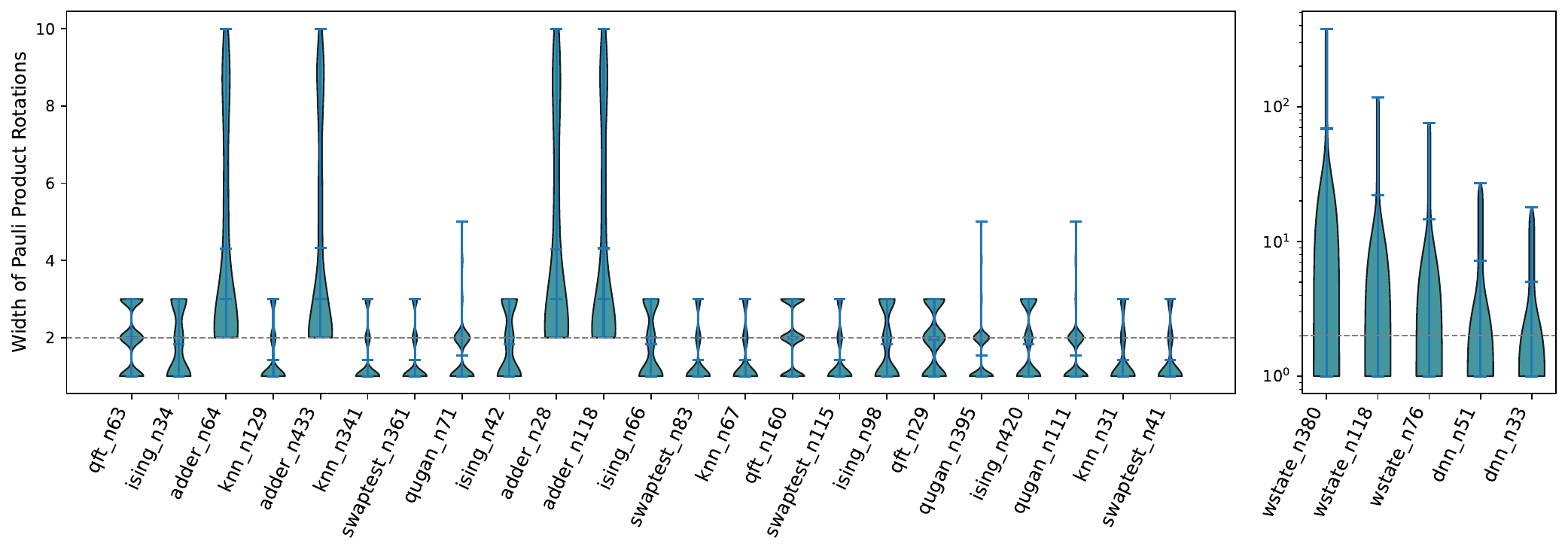}
    \caption{\revision{Distribution of PPRs length across
        representative QASMBench~\cite{10.1145/3550488} Benchmarks.
        Notably, all benchmark contains non-trivial interactions of width
        $> 2$. On the right we separated out benchmarks with larger with
    rotations and used logscale.}}
    \label{fig:violin-all}
  \end{subfigure}
  \begin{subfigure}{\textwidth}
    \centering
    \includegraphics[width=\linewidth]{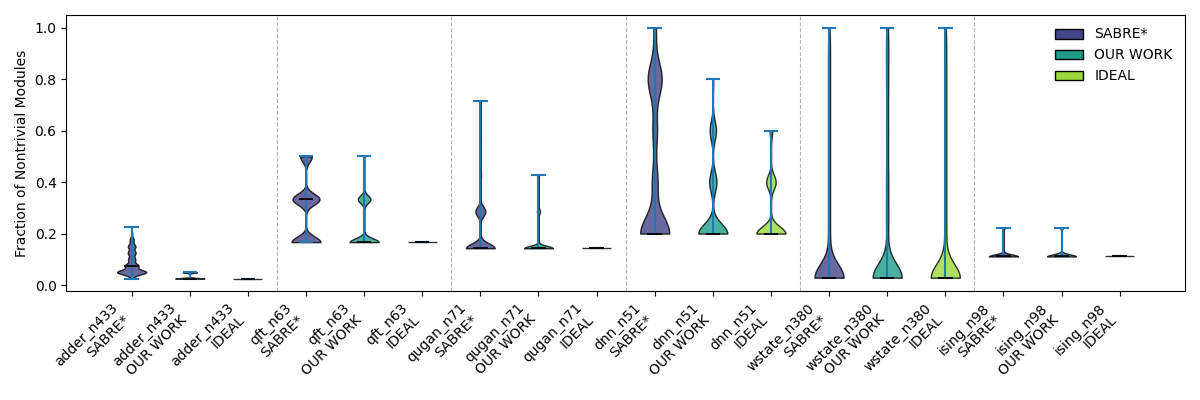}
    \caption{\revision{Comparing the fraction of modules used for a
        Pauli Product Rotation after partitioning, between our mapping
        policy, SABRE* and the ideal case (i.e., $\lceil\frac{n}{11}
    \rceil$, where $n$ qubits are involved in the rotation).}}
    \label{fig:violin-proof}
  \end{subfigure}
  \caption{\revision{Comparing the width of the Pauli Product
  Rotations before and after partitioning.}}
\end{figure*}

\subsubsection{Creating Trivial In-module Measurements}
\label{label:trivial-intermodule}
A given rotation may have a span of 12 bits $i_1... i_{12}$ of all
identities inside a single module, producing a trivial in-module rotation.
This avoids both in-module and inter-module measurements.
Mappings that align identity blocks with module boundaries can
therefore reduce inter-module operations across all rotations.

\subsubsection{Finding Efficient Paths to the Factory}
\label{label:efficient-paths}
For each rotation, a T-state is generated at the factory and must
propagate to the required modules using inter-module measurements.
In a linear topology (see Figure~\ref{fig:intro-fig-a}), this
propagation requires the T-state to traverse each intermediate module
sequentially to reach the farthest target module.
However, by leveraging trivial in-module measurements
(Section~\ref{label:trivial-intermodule}), a rotation may involve
anywhere from $1$ module to $M$ modules, rather than always involving
all $M$ modules.
Thus, the physical placement of modules relative to the factory
directly impacts the number of inter-module measurements, and
therefore the overall \revision{program error rate}.
Consequently, the inter-module measurement cost for a rotation is
equal to the number of edges in the minimum spanning tree that
connects the factory and all modules participating in that rotation.

\section{Our Proposal: Enhanced Logical Mapping}
Using the insights developed in Section~\ref{sec:reduce-intermodule},
we aim to reduce the \textit{inter-module measurements} using the two
methods discussed in Sections~\ref{label:trivial-intermodule}
and~\ref{label:efficient-paths}.
We find that each of these problems requires tailored heuristics to
solve the respective clustering and assignment problems individually.
We define the problem of reducing nontrivial inter-module
measurements and finding efficient paths to the factory as the
\textit{clustering} and the \textit{assignment} problem, respectively.
As a result, we propose a two step approach that solves both of these problems.
We first focus on reducing inter-module measurements by using a
hypergraph partitioning to re-partition the circuit by constructing
clusters that minimize the number of modules involved in each
rotation, thereby reducing the number of non-trivial modules involved
in each rotation.
Secondly, we use the partitioned clusters to make assignments to
physical locations based on the ability to recycle repeated paths of
certain sets of module(s), thereby minimizing the total inter-module
cost for the chosen clustering.

\begin{figure*}
  \centering
  \includegraphics[width=\linewidth]{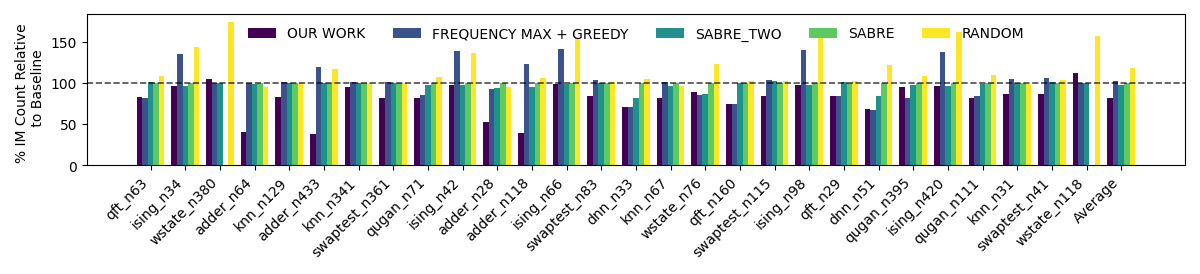}
  \caption{Relative reduction in \revision{the number of inter-module
  measurements} achieved as percent improvement over SABRE.}
  \label{fig:intermodule-reduction}
\end{figure*}

\subsection{Efficient Clustering via Hypergraph Partitioning}
To construct clusters of logical qubits that should be grouped within
a module, we represent each interaction in a given rotation as an
edge in a hypergraph, where an edge can span more than 2 nodes. We
assign weights to these hyperedges using the frequecy of the
rotation, i.e., if a certain rotation is executed multiple times in
any permutation.
This is essential, as Figure~\ref{fig:violin-all} shows that every
benchmark has rotations of width $> 2$.
The optimization problem is to then partition the graph into $M$ cuts
such that the weight across each of the cuts are minimized, creating
clusters of size up to $11$, where qubits that interact frequently
are in the same group.

\begin{figure}[h]
  \centering
  \includegraphics[width=\linewidth]{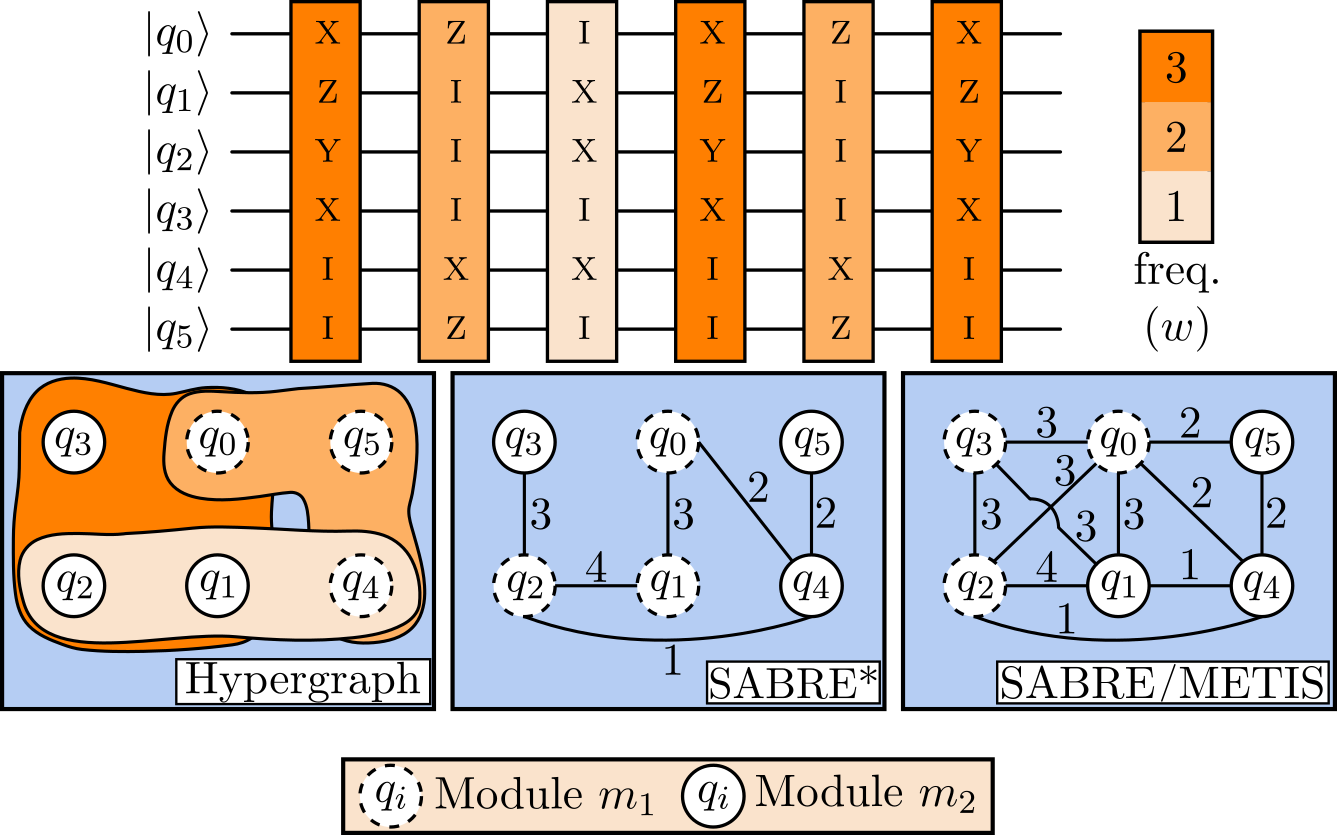}
  \caption{Example of a PBC circuit partitioned using four methods:
    hypergraph (left),  SABRE (middle), and SABRE*/METIS (right). Each
    Pauli-layer involves nontrivial operations on multiple qubits
    (frequency of repetition are indicated by the color gradient). Each
    method builds an interaction graph \revision{weighted by} the
    number of interactions each qubit has with each other. The
    hypergraph appends a \revision{frequency-weighted (hyper)edge} for
    each operation, SABRE* adds linear edges per operation, and
    SABRE/METIS add a quadratic number of edges for each operation.
  Qubits are then \revision{partitioned into modules by computing min-cuts}.}
  \label{fig:mappers}
\end{figure}

We use KaHyPar\cite{KaHyPar1,KaHyPar2} with $\varepsilon$ value at
0.06 to implement this partitioning.
Creating clusters using KaHyPar reduces the number of modules
involved in each rotation, allowing for more possible trivial
inter-module measurements. We compare against SABRE, SABRE*, as well
as METIS in Figure~\ref{fig:mappers} wherein we show a toy example
such that the hypergraph partitioner generates a partition that
reduces the inter-module cost. Across all benchmarks, we report the
reduction in number of non-trivial inter-module measurements
outputted by KaHyPar in Figure~\ref{fig:intermodule-reduction}.

\revision{Figure \ref{fig:violin-proof} also shows how our mapper is
  able to minimize the number of nontrivial rotations, shifting the
distribution of nontrivial modules on a per-rotation basis.} We find
that, apart from \revision{wstate and ising}, KaHyPar is able to
reduce the \revision{number} of inter-module measurements compared to
\revision{SABRE*}. \revision{These benchmarks offer no scope for
  efficient partitioning since both of them have all-to-all qubit
  interactions, wstate in the form of large weight interactions that
  span all modules, and ising in the form of all qubit combinations of
  low-weight interactions. This is also reflected in
  Figure~\ref{fig:intermodule-reduction}, where the improvement from
our mapper is significant on all benchmarks classes except these two.}
Furthermore, it performs best on benchmarks which have rotations with
higher density of widths that are $> 2$, but do not span the entire
width of the program.
Overall, Figure~\ref{fig:mainresults} demonstrates that
hypergraph-partitioning–based clustering strategies, when coupled
with any of the assignment policies, consistently reduce the resulting error.

\subsection{A Priority Based Algorithm for Efficient
Assignment}\label{sec:priority-assignment}

\begin{algorithm}
  \scriptsize
  \caption{Priority Assignment Heuristic}
  \label{algorithm:priority-heuristic}
  \begin{algorithmic}[1]
    \Procedure{AssignPriorities}{$\mathsf{Rotations}, \mathsf{numModules}$}
    \State{$\mathsf{Priorities} \gets \{\}$}
    \State{priority $\gets 0$}
    \State{modules $\gets \{(\text{id: modID}, \text{freq: }0,
        \text{rotations: }\emptyset)\mid \text{modID} \in [1,
    \mathsf{numModules}]\}$}
    \Statex{\Comment{initialize a list of modules with frequencies}}
    \ForAll{rotation $\in\mathsf{Rotations}$}
    \ForAll{module $\in$ rotation}
    \State{modules[module].freq $\gets$ freq $+$ rotation.freq}
    \State{modules[module].rotations $\gets$
    modules[module].rotations $\cup$ rotation}
    \Statex{\Comment{update the frequency of \emph{module} and add
    \emph{rotation}}}
    \EndFor{}
    \EndFor{}
    \While{modules $\neq \emptyset$}
    \State{module $\gets$ \Call{GetModuleWithLeastFreq}{modules}}
    \State{modules $\gets$ \Call{RemoveModule}{modules, module}}
    \State{modules $\gets$ \Call{DelRotationAndFrequencies}{modules,
    module.rotations}}
    \Statex{\Comment{delete the rotations and frequencies associated
    with \emph{module}}}
    \State{$\mathsf{Priorities} \gets \mathsf{Priorities} \cup
    (\text{module.id, priority})$}
    \State{priority $\gets$ priority $+ 1$}
    \EndWhile{}
    \State{\Return{$\mathsf{Priorities}$}}\Comment{returns a
    dictionary of $(\mathsf{moduleID}, \mathsf{priority})$ pairs}
    \EndProcedure{}
  \end{algorithmic}
\end{algorithm}

\begin{figure}[htbp!]
  \centering
  \includegraphics[width=0.6\linewidth]{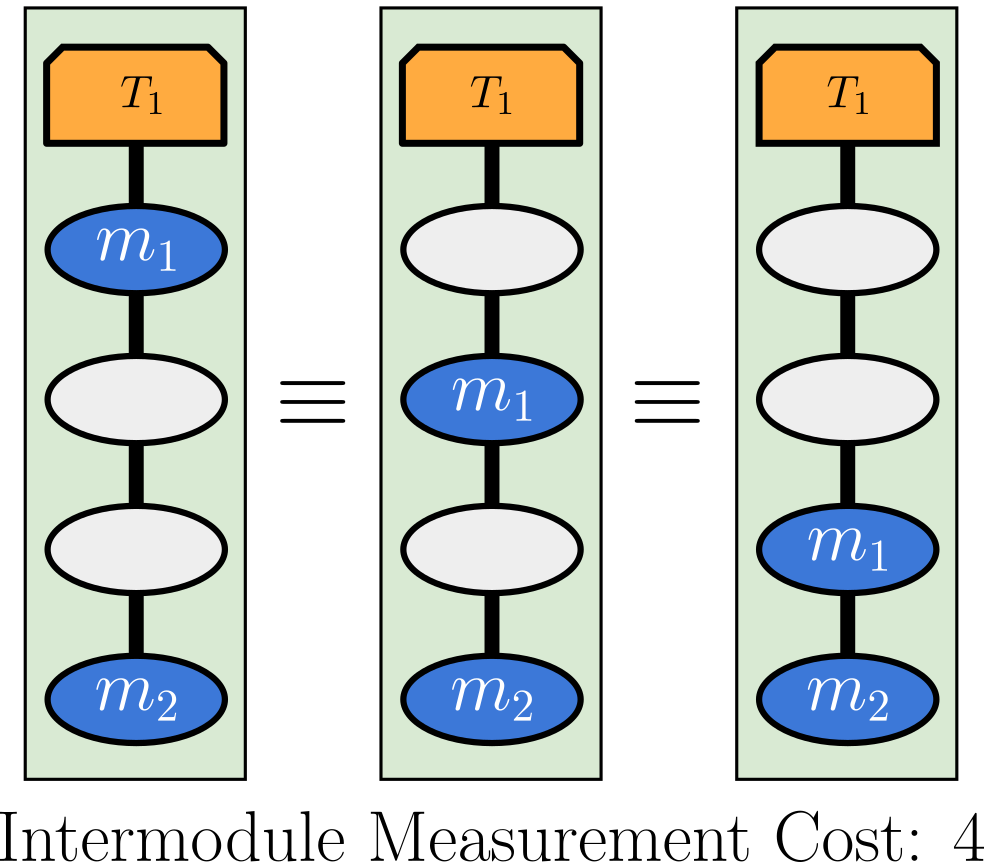}
  \caption{Regardless of module placement for $m_1$ the inter-module
    cost is always 4, as $m_2$ was placed 4 modules away, ultimately
    requiring 4 inter-module connections between the magic state
  factory and $m_2$.}
  \label{fig:module_cost_equiv}
\end{figure}

\begin{figure*}
  \centering
  \revisionbox{\includegraphics[width=\linewidth]{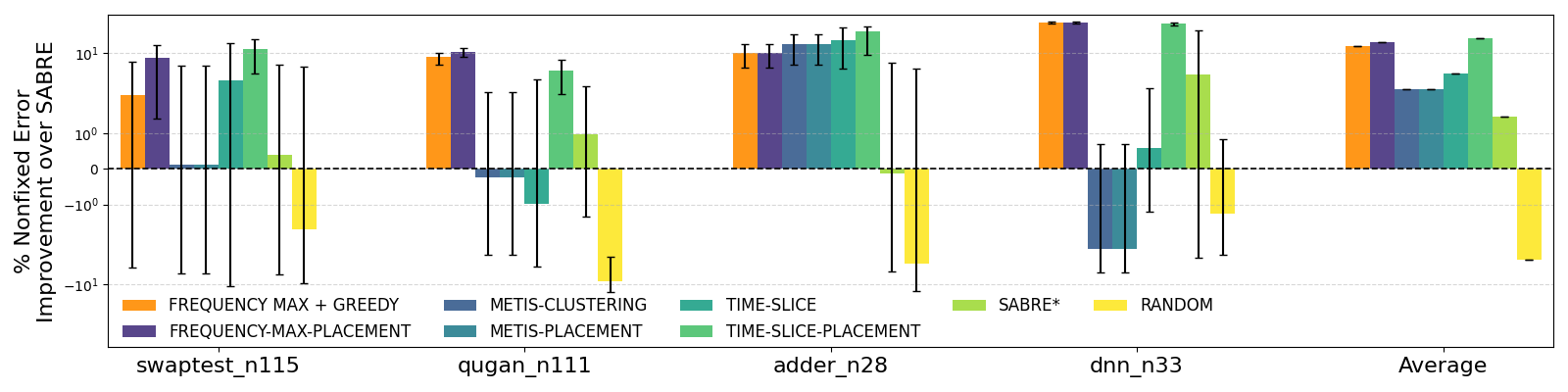}}
  \caption{\revision{We demonstrate program error rate improvements
      obtained from using Algorithm~\ref{algorithm:priority-heuristic} in
      conjunction with different partitioning policies.
      Algorithm~\ref{algorithm:priority-heuristic} enhances the
      improvements obtained by each partitioning strategy. Note that the
      average reported is over the depicted subset of benchmarks and not
      the entire benchmark suite used in
  Section~\ref{sec:evaluation}.}}\label{fig:placement_improvements}
\end{figure*}

Finding efficient paths to the factory must leverage the ability to
assign module locations such that common paths are reused. Moreover,
if a module is involved in fewer rotations, it will likely have a
smaller contribution to the inter-module measurement cost, and thus
the overall error rate. Such modules should be placed further away
from the factory. Inspired from these observations, we propose the
priority heuristic described in in
Algorithm~\ref{algorithm:priority-heuristic}. One of the key insights
of our heuristic lies in the observation from
Figure~\ref{fig:module_cost_equiv}, wherein we note that for a linear
architecture, the cost of the rotation is solely determined by the
placement of the non-trivial module that is the furthest away from
the factory. This insight holds true even for a grid architecture
with a limited number of factories. Routing to modules on grids with
a constant width involves limited interactions along the rows and
most of the inter-module cost comes from routing along the columns.
We detail this discussion in Section~\ref{sec:grid-routing}.
\par Our heuristic counts the number of rotations each module is
involved in, and starts assigning priorities outside in, that is, it
first identified modules with the least frequencies and assigns it
the least possible priority. It then removes all rotations that are
incident on this module, We also update the frequencies of the other
modules that are involved in the removed rotations, since the
remaining unassigned modules do not affect the inter-module cost of
such rotations anymore. We repeat the above algorithm until there are
no more modules left to be assigned.
\par We compare the difference between hypergraph partitioning with a
greedy assignment heuristic versus our priority heuristic to better
see an isolation of variables. We also compare all policies against
each other and we report the performance across all benchmarks in
Figures~\ref{fig:mainresults}. We also show the improvements gained
by implementing Algorithm~\ref{algorithm:priority-heuristic} on top
of baseline mapping strategies in Figure~\ref{fig:placement_improvements}.

\subsubsection{\revision{Optimality of
Algorithm~\ref{algorithm:priority-heuristic}}}
\revision{We prove the optimality of the placement heuristic for the
  case of a line architecture and a single factory at the end of the
  line. The same proof idea extends to a grid architecture and the
  multiple factories case. For simplicity, we assume that all modules
  have distinct frequencies as computed in
  Algorithm~\ref{algorithm:priority-heuristic}. This guarantees that
  the optimal priority assignment is unique. Let the optimal priority
  assignment be denoted by $\mathcal{M}^*$. Furthermore, let the
  assignment returned by Algorithm~\ref{algorithm:priority-heuristic}
  be denoted by $\mathcal{M}$, and assume that $\mathcal{M} \neq
  \mathcal{M}^*$. We will now show that this leads to a contradiction,
  and therefore, the assignment given by
  Algorithm~\ref{algorithm:priority-heuristic} is optimal and equal to
$\mathcal{M}^*$.}
\revision{Since we assumed $\mathcal{M} \neq \mathcal{M}^*$, there
  exists some priority on which $\mathcal{M}$ and $\mathcal{M}^*$
  differ. Consider the smallest such priority $i$, and let the module
  assignment by $\mathcal{M}, \mathcal{M}^*$ at priority $i$ be $m_i,
  m_i^*$ respectively, where $m_i \neq m_i^*$. Since we assumed all
  modules have distinct frequencies, and
  Algorithm~\ref{algorithm:priority-heuristic} chooses module $m_i$
  such that it has the least frequency, we have $\text{F}(m_i^*) >
  \text{F}(m_i)$, where $\text{F}(\cdot)$ is the frequency function.
  Therefore, the cost of inter-module rotations after the
  $i\textsuperscript{th}$ assignment can be computed as (using the
  approach from Figure~\ref{fig:module_cost_equiv}),
  \begin{align*}
    \text{cost}_i(\mathcal{M}) &= \text{cost}_{i-1}(\mathcal{M}) +
    \text{F}(m_i) \cdot (n_\text{m} - i)\\
    &= \text{cost}_{i-1}(\mathcal{M}^*) + \text{F}(m_i) \cdot (n_\text{m} - i)\\
    &< \text{cost}_{i-1}(\mathcal{M}^*) + \text{F}(m_i^*) \cdot
    (n_\text{m} - i) < \text{cost}_i(\mathcal{M}^*),
  \end{align*}
  where in the second step we used the fact that the assignments were
  identical up to step $i-1$, and $n_\text{m}$ is the number of
  modules. Since we have that
  Algorithm~\ref{algorithm:priority-heuristic} achieves a lower cost
  than the optimal assignment, it could not have been the case that
  $m_i$ differs from $m_i^*$, which leads to a contradiction. %
}

\begin{figure}
  \centering
  \includegraphics[width=\linewidth]{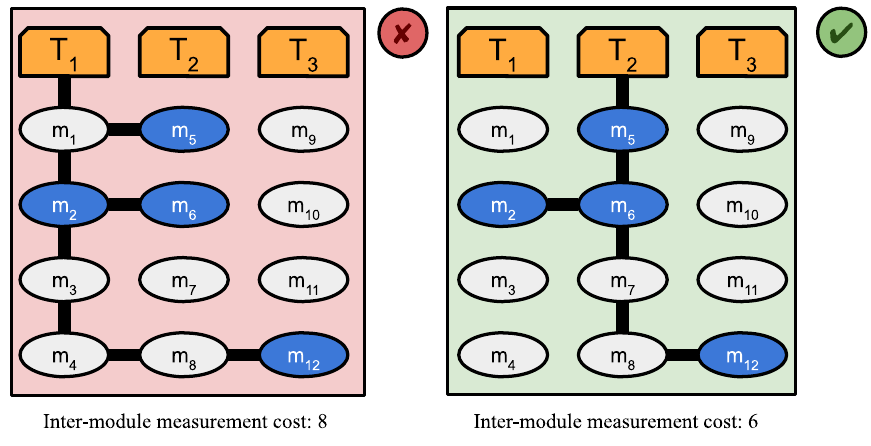}
  \caption{An example of the path-finding algorithm on the long-grid
    topology, where three T-state factories $T_1, T_2, T_3$ are
  positioned along the short edge of the grid.}
  \label{fig:factory-grid-explain}
\end{figure}

\subsection{Routing on Grid Topologies}\label{sec:grid-routing}

Although the tour de gross architecture assumes a linear connection
topology between the modules and the factory, we also evaluate our
mapper on a more realistic topology that researchers are currently
investigating for fault-tolerant architectures~\cite{IBM2025LargeScaleFTQC}.
Since the circuit execution in the tour de gross ISA is largely
sequential, and each module consists of $11$ logical data qubits, we
only have $40$ modules for the largest benchmark. In addition
\revision{to} this, the factories are no longer the bottlenecks in
terms of execution times (Table~\ref{tab:error-rates}) of the gates.
Therefore, we consider factories that are arranged along the short
edge of a \textit{long-grid}, where the grid’s length is
significantly larger than its width. For a grid with $M$ modules and
$f$ factories, where $f \ll M$, we can define the dimensions of the
grid as $\lceil M/f\rceil \times f$ and treat $f$ as a constant.

Even on such \textit{long-grids}, brute-forcing the optimal path that
minimizes the inter-module cost requires searching an exponential
space. Therefore, we propose a fast path-finding appraoch that is
nearly optimal for such topologies.
We find that since the grid is much longer than it is wide, and
T-state must originate from a single factory, it is more effective to
explore depth-wise along the long dimension from the central factory
and branch out horizontally only when necessary.
We  demonstrate this in Figure~\ref{fig:factory-grid-explain}, where
we explore length-wise to the maximum required depth while ensuring
that the horizontal (width-wise) movement incurs minimal inter-module
measurements, while still achieving the optimal depth-wise cost.
Since we have a constant number of factories in a \textit{long-grid},
we can perform this search for each possible factory as a starting
point and choose the one which minimizes the inter-module cost.

\section{Evaluation and Methodology}\label{sec:evaluation}
\begin{figure*}
  \centering
  \begin{subfigure}{\textwidth}
    \includegraphics[width=\linewidth]{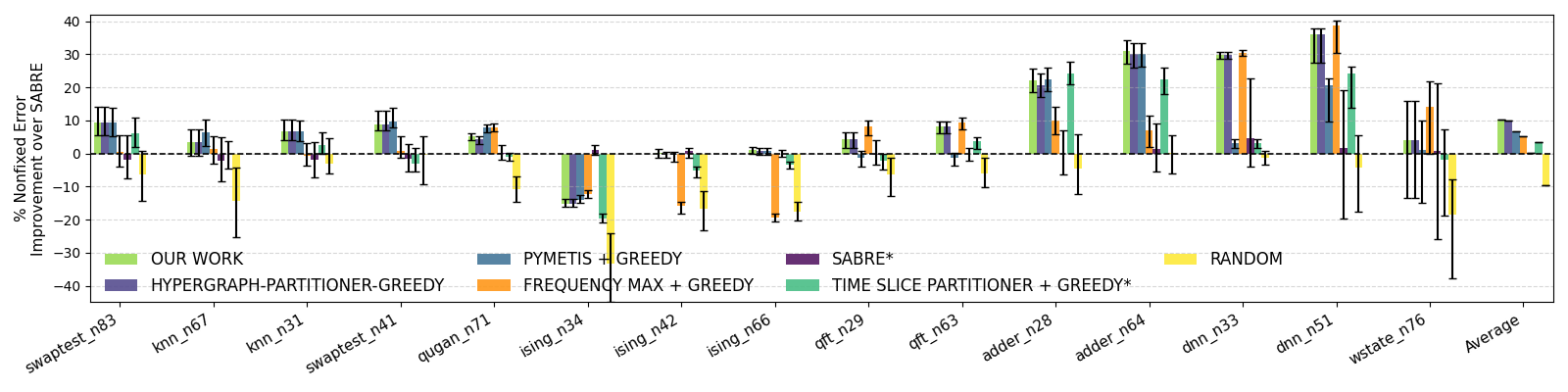}
  \end{subfigure}
  \begin{subfigure}{\textwidth}
    \includegraphics[width=\linewidth]{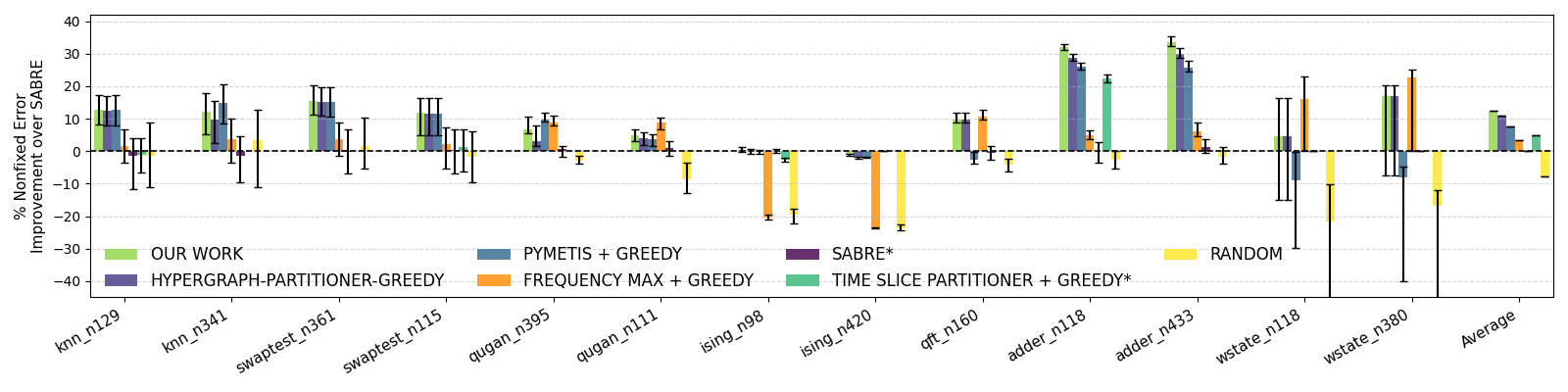}
  \end{subfigure}
  \caption{Results for medium-sized \revision{(top)} and large-sized
    benchmarks \revision{(bottom)} from the QASMBENCH
    suite\cite{10.1145/3550488}, showing relative percentage
    improvements in program error rates of our mapper over baseline compared
    to SABRE, SABRE*, Time Slice Partitioning*, and METIS on the
  non-fixed/optimizable portion of the compilation.}\label{fig:mainresults}
\end{figure*}

\subsection{Generalized Method}
\label{sec:general-eval}
We first compile benchmarks from QASMbench~\cite{10.1145/3550488} to
obtain our logical circuit benchmarks.
These circuits are then translated to a Pauli-based computing model
following the commutation rules discussed
in~\cite{litinski2019gameofsurfacecodes,tourdegross}
(Figures~\ref{fig:PBCmeasurement},~\ref{fig:TDGarchitecture}), such
that each circuit is represented as a list of PPR.
Given a benchmark of $n$-logical qubits, the number of Gross-code
modules is $M = \lceil\frac{n}{11}\rceil$.
Each mapping policy takes the rotation list and the number of modules
as input, and outputs a mapping from the logical qubit to the module.
Once a mapping is produced, we decompose each rotation into a set of
11-qubit Pauli-blocks at each timestep.
Each rotation, $P(\varphi) = (P_{m_1}\otimes  P_{m_2} \otimes \cdots
P_{m_M})(\varphi)$, is annotated with a set of modules, $\mathcal{M}
= \{m_i\mid P_i \neq I^{\otimes 11}\}$ which represents the
non-trivial modules involved in that rotation.
\revision{In the case of a linear topology, the optimal path is the
  path that connects module $\max(\mathcal{M})$ to the factory, i.e.,
the module that is the furthest away from the factory}. Using this
path, we calculate the number of inter-module measurements needed for
a given rotation, and we use the table provided in~\cite{tourdegross}
to get the in-module measurement cost upon each of the
$\left|\mathcal{M}\right|$ modules participating in a nontrivial rotation.
\revision{The total program failure rate ($P_\text{total}$) is
  calculated based on the complement of the success probability:
  \[P_\text{total} = 1 - \prod_{\text{operation}_i \in L} (1 -
  p_\text{operation})\]
  where L denotes all the bicycle instructions in the circuit, and
  $p_\text{operation}$ is obtained from Table~\ref{tab:error-rates}.
  Note that the total program failure rate corresponds directly to
  application performance. In practice, the confidence level for an
  application is given by the product of the number of shots of the
  application and the success probability. We can expect an application
  to be run with $\text{\# shots} \cdot{(1-P_\text{total})^{-1}}
  \approx \text{\# shots}\cdot(1+P_\text{total})$ (this approximation
  holds in the FTQC regime where failure rates are small). Therefore,
  an improvement of $c\%$ in total program failure rate corresponds to
  a $c\%$ improvement in application wall time, or, a $c\%$ improvement
on the depth of a given circuit for the same execution time.}
We assume the physical error rates of all bicycle instruction to be
$p = 10^{-4}$.
The synthesis cost is calculated with a precision of $10^{-4}$,
giving $\sim40$ T gates (Cliffords are commuted to the end of the circuit).
Each synthesized T gate requires an additional inter-module
measurement with the module adjacent to the factory (as shown in
Figure~\ref{fig:TDGarchitecture}).

\subsection{Baselines}
Since the logical qubit mapping for Bivariate Bicycle (BB) codes
lacks previous standardized methodologies, we evaluate five different
baseline strategies throughout the paper.
Some of these, such as SABRE~\cite{SABRE},
time-slice~\cite{10.1145/3387902.3392617}, and
METIS~\cite{autobraid}, are adapted from NISQ-based and FTQC-based
mappers, and the remaining baselines rely on naive heuristics
designed specifically for the logical-module setting.
For each baseline we consider both the \textit{clustering} heuristic
to partition qubits into modules, and the \textit{assignment}
heuristic to assign the modules to a physical location.
Due to nondeterminism in several baselines, we perform 10 independent
runs and report mean values with error bars. \revision{We also report
  the mean and standard deviation for the compilation time taken by
  each mapping policy in Table~\ref{tab:runtime_stats}. These were
computed on a Lenovo machine with Intel Ultra 9 185H and 64 GB of RAM.}

\begin{table}[t]
  \setlength\tabcolsep{0pt}
  \centering
  \begin{tabular*}{\columnwidth}{@{\extracolsep{\fill}}lccc}
    \hline
    \textbf{Policy} & \textbf{Mean (s)} & \textbf{Std (s)} &
    \textbf{\#Timeouts} \\
    \hline
    RANDOM & 1.71 & 3.45 & 0 \\
    FREQUENCY MAX + GREEDY & 2.11 & 4.25 & 0 \\
    Our Mapping Policy & 2.61 & 5.54 & 0 \\
    HYPERGRAPH-PARTITIONER-GREEDY & 2.62 & 5.31 & 0 \\
    METIS-CLUSTERING & 3.23 & 5.89 & 0 \\
    SABRE* & 4.37 & 8.93 & 0 \\
    SABRE & 5.19 & 11.89 & 2 \\
    TIME-SLICE & 75.14 & 93.55 & 10 \\
    \hline
  \end{tabular*}
  \caption{\revision{Runtime statistics (mean and standard deviation)
      across benchmarks sorted by mean. Timeouts correspond to runs that
  took longer than 5 minutes.}}\label{tab:runtime_stats}
\end{table}

\subsubsection{NISQ-Based and FTQC-Based Baselines}

\begin{figure*}[ht!]
  \centering
  \revisionbox{\includegraphics[width=\linewidth]{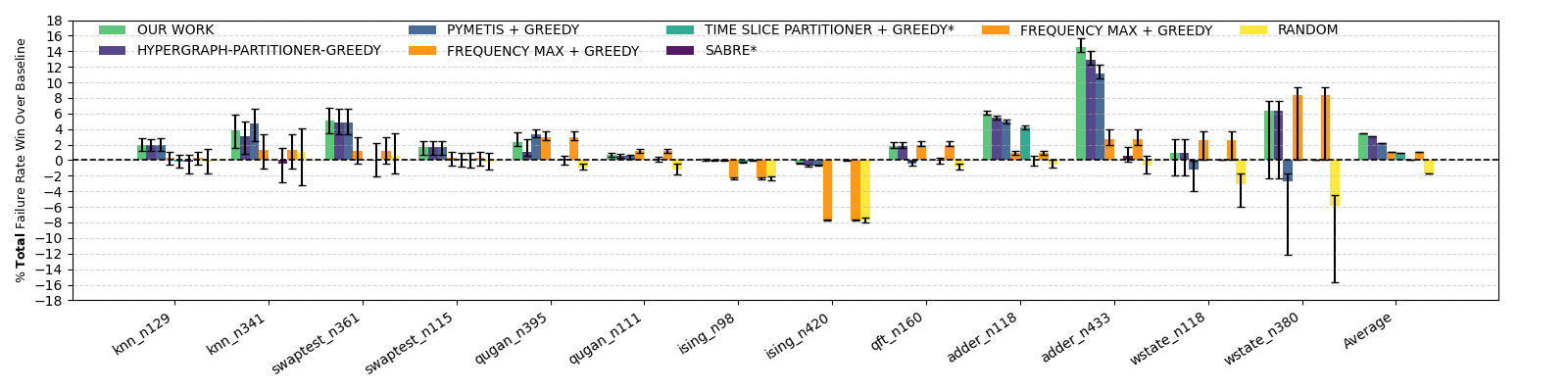}}
  \caption{Including the fixed T-synthesis error, we report the total
    \revision{program failure rate} expressed as a percentage relative
    to the baseline, SABRE. Relative percentage improvement over SABRE
  is shown for each mapper.}
  \label{fig:total-ler-plot}
\end{figure*}

\begin{figure*}[ht!]
  \centering
  \begin{subfigure}{0.325\textwidth}
    \centering
    \revisionbox{\includegraphics[width=\linewidth]{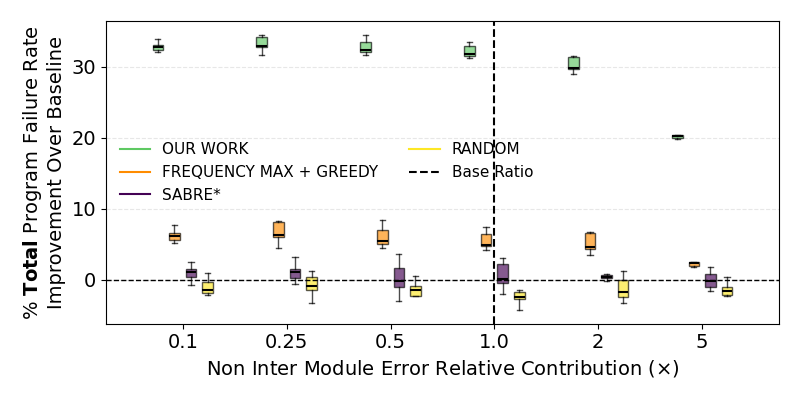}}
    \caption{\revision{adder\_433 benchmark}}\label{fig:sensitivity_a_adder}
  \end{subfigure}
  \begin{subfigure}{0.325\textwidth}
    \centering
    \revisionbox{\includegraphics[width=\linewidth]{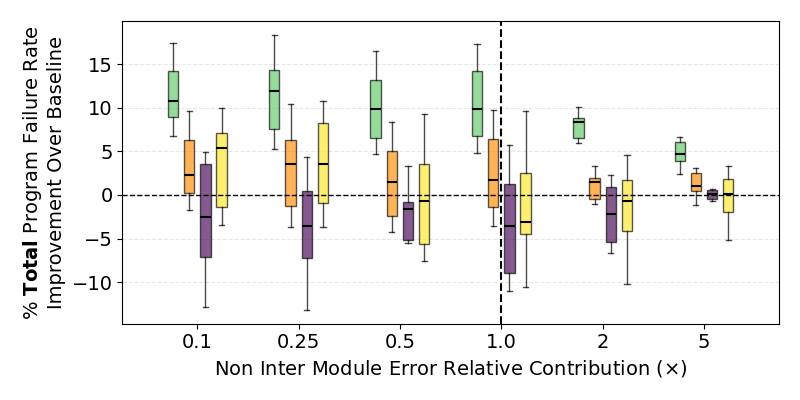}}
    \caption{\revision{knn\_341 benchmark}}\label{fig:sensitivity_a_knn}
  \end{subfigure}
  \begin{subfigure}{0.325\textwidth}
    \centering
    \revisionbox{\includegraphics[width=\linewidth]{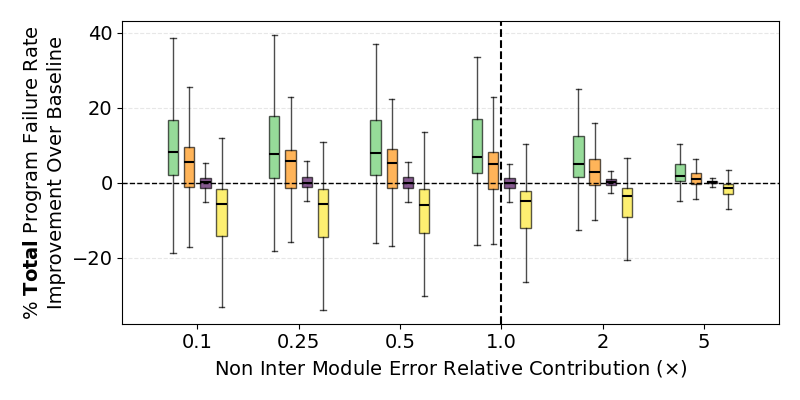}}
    \caption{\revision{Averaged across all
    benchmarks}}\label{fig:sensitivity_a_avg}
  \end{subfigure}
  \caption{\revision{Sensitivity to the Rz Synthesis Error
      contribution (note that this plot assumes $P_C = 0$ only for
        injections since injection error rates have not been determined in
  literature for arbitrary extractor architecture).}}\label{fig:sensitivity_a}
\end{figure*}

Four of our baselines are inspired from state of the art NISQ
mappers, which we adapted for the bicycle code logical qubit mapping
problem: SABRE, SABRE*~\cite{SABRE}, Time Slice
Partitioning*~\cite{10.1145/3387902.3392617} and METIS~\cite{autobraid}.
Since the NISQ baselines do not consider $k$-qubit interactions (for
$k > 2$), we model the rotations as 2 versions of each NISQ baseline:
one version where CXs are all to all connected between a length $k$
interaction, and one version where a single CX connects the length
$k$ interaction in a single linear chain. On programs with $n$
qubits, the first approach generates graphs with $O(n^2)$ edges,
whereas the latter approach generates graphs with only $O(n)$ edges.
However, the first approach has the advantage of capturing all
possible orderings of qubits, whereas the second approach is faster.
Indeed we see that the all-to-all variants exhibit severe
computational overhead and frequently encounter timeouts or memory exhaustion.
For this reason, SABRE has missing data points for certain
benchmarks, and the all-to-all variant of Time Slice Partitioning was
discarded entirely. We denote the latter version of each baseline with a star.
Even Time Slice Partitioning* occasionally fails to complete on the
linear chain connectivity.

Due to this inconsistency, SABRE serves as our \textit{primary NISQ
baseline}, with SABRE* used only when SABRE itself times out.
We pass to SABRE a line connectivity graph of all logical qubits,
since SABRE does not have the ability to consider modules/module
connectivity, and we assume that all qubits $i$, where
$\lceil\frac{i}{11}\rceil = k$ are grouped together into module $m_k$
from the output line to artificially make the clusters.
Time Slice Partitioning in contrast, can explicitly model cluster
structure, given the device graph and the number of partitions
$\lceil \frac{n}{11}\rceil$ of size at most 11.
However, it does not address the problem of relative orderings of these modules.
We therefore use a greedy assignment policy, placing the most
frequently used module (as determined by the partition policy)
closest to the factory.

\subsubsection{Customized Greedy Baseline}

We also use our own naive heuristic to serve as a baseline that was
purposed for the logical qubit mapping problem, we call ``frequency
maximizing" clustering.
This counts the frequencies of each logical qubit being involved
non-trivially in a rotation, and creates a list in descending order
from most used to least used logical qubits.
The policy then \revision{assigns qubit $q_i$ at index $i$ in the
list into module $m_k$, where $k = \lceil\frac{i}{11}\rceil$}.
We combine this with the same greedy assignment heuristic, as in the
case of Time Slice Partitioning policy, where the most used modules
are physically assigned to locations closest to the factory.

\subsection{Evaluation on Non-fixed Error vs Total Error}

\revision{Referring to the analysis in
  Section~\ref{sec:reduce-intermodule}, placement and routing cannot
  address the overheads associated with the $Rz$ state synthesis and
  the ensuing inter-module measurements required to inject T-states
  into the module for executing rotations.
  Additionally, these operations take up a majority of the total
  failure probability ($P_\text{error}$ as defined in
  Section~\ref{sec:general-eval}), and cannot be altered by using
  different mapping policies. As a result, we also compare the
  contribution to the total failure probability from the in-module and
  inter-module operations, i.e., the mapping-dependent failure
  probability (which we call the non-fixed error, or
  $P_\text{non-fixed}$). We report the errors for $P_\text{non-fixed}$
  in Figure~\ref{fig:mainresults} and the errors for $P_\text{total}$
in Figure~\ref{fig:total-ler-plot}.}

\subsection{\revision{Sensitivity Analysis}}

\subsubsection{\revision{Sensitivity to the Rz Synthesis Error Contribution}}
\begin{figure*}[ht!]
  \centering
  \begin{subfigure}{0.325\textwidth}
    \centering
    \revisionbox{\includegraphics[width=\linewidth]{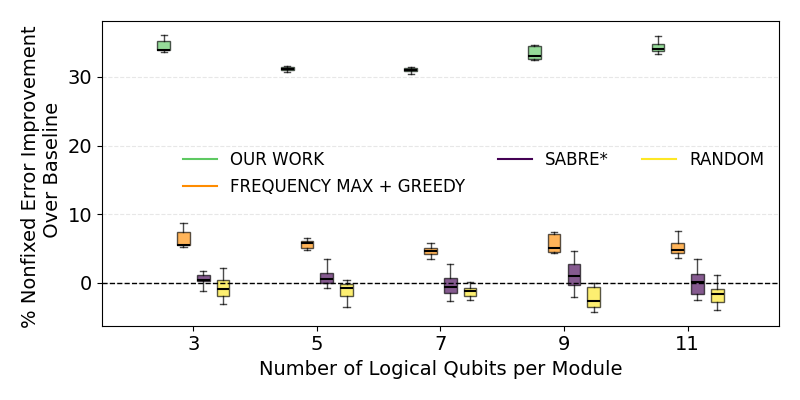}}
    \caption{\revision{adder\_433 benchmark}}\label{fig:sensitivity_b_adder}
  \end{subfigure}
  \begin{subfigure}{0.325\textwidth}
    \centering
    \revisionbox{\includegraphics[width=\linewidth]{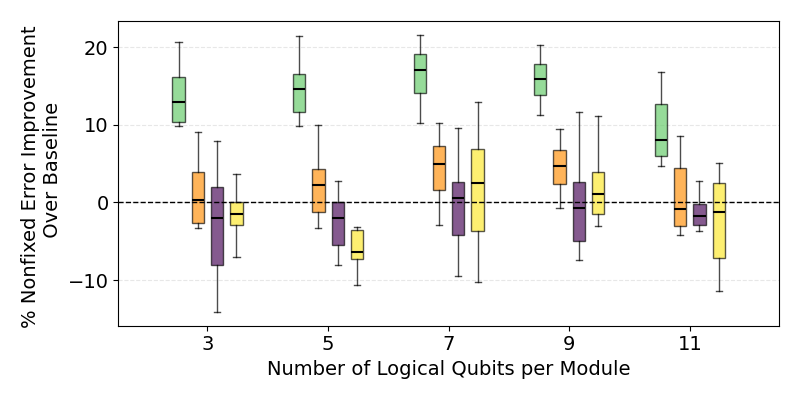}}
    \caption{\revision{knn\_341 benchmark}}\label{fig:sensitivity_b_knn}
  \end{subfigure}
  \begin{subfigure}{0.325\textwidth}
    \centering
    \revisionbox{\includegraphics[width=\linewidth]{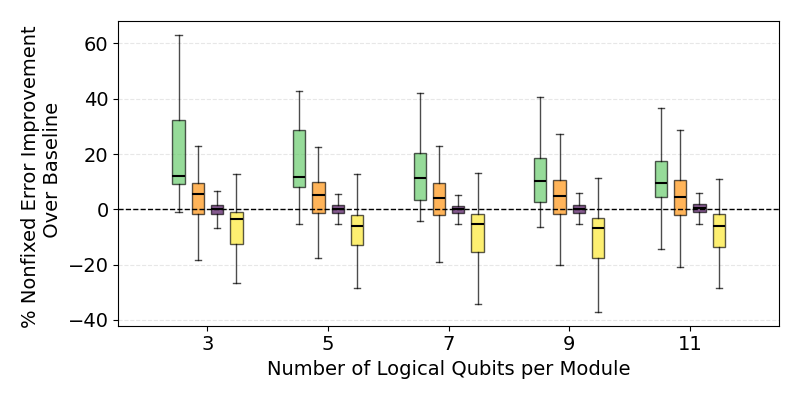}}
    \caption{\revision{Averaged across all
    benchmarks}}\label{fig:sensitivity_b_avg}
  \end{subfigure}
  \caption{{Exploring different amounts of logical qubit numbers per
      module and the effect on non-fixed error
  improvement.}}\label{fig:sensitivity_b}
\end{figure*}

\begin{figure*}[ht!]
  \centering
  \begin{subfigure}{0.325\textwidth}
    \centering
    \revisionbox{\includegraphics[width=\linewidth]{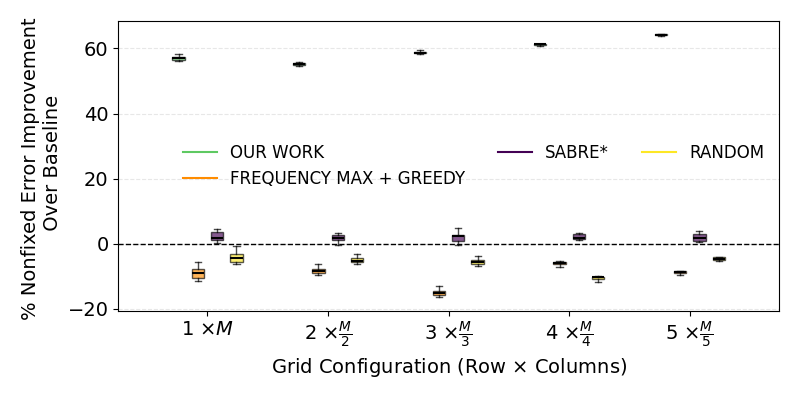}}
    \caption{\revision{adder\_433 benchmark}}\label{fig:sensitivity_d_adder}
  \end{subfigure}
  \begin{subfigure}{0.325\textwidth}
    \centering
    \revisionbox{\includegraphics[width=\linewidth]{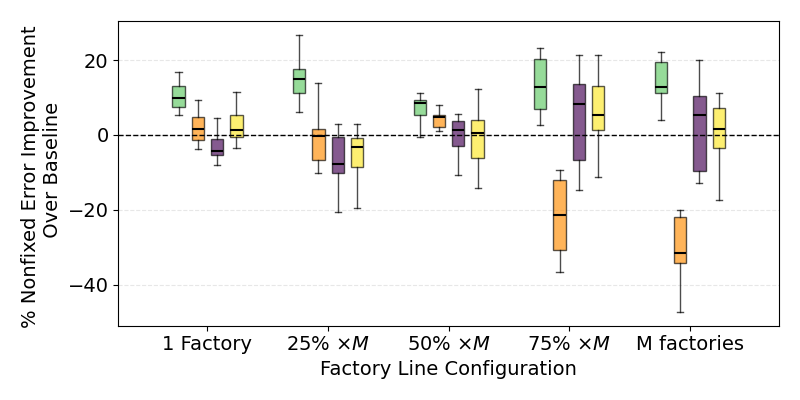}}
    \caption{\revision{knn\_341 benchmark}}\label{fig:sensitivity_d_knn}
  \end{subfigure}
  \begin{subfigure}{0.325\textwidth}
    \centering
    \revisionbox{\includegraphics[width=\linewidth]{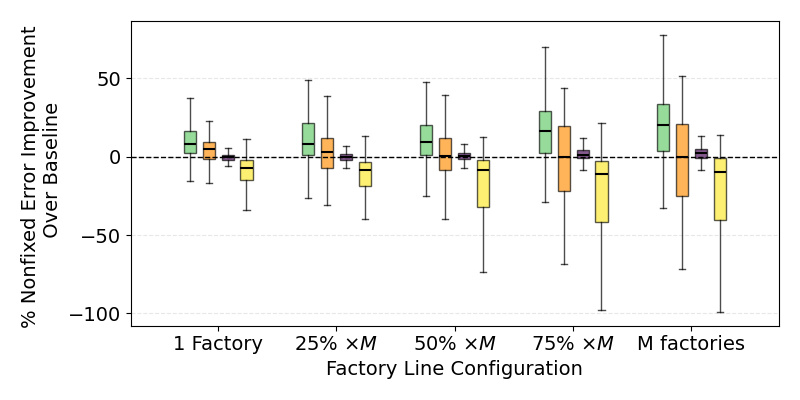}}
    \caption{\revision{Averaged across all
    benchmarks}}\label{fig:sensitivity_d_avg}
  \end{subfigure}
  \caption{{Factory density as a function of the number of modules,
      $M$ where a factory is added at every $\lceil x\% \times M \rceil$
  spots.}}\label{fig:sensitivity_d}
\end{figure*}

\begin{figure*}[ht!]
  \centering
  \begin{subfigure}{0.325\textwidth}
    \centering
    \revisionbox{\includegraphics[width=\linewidth]{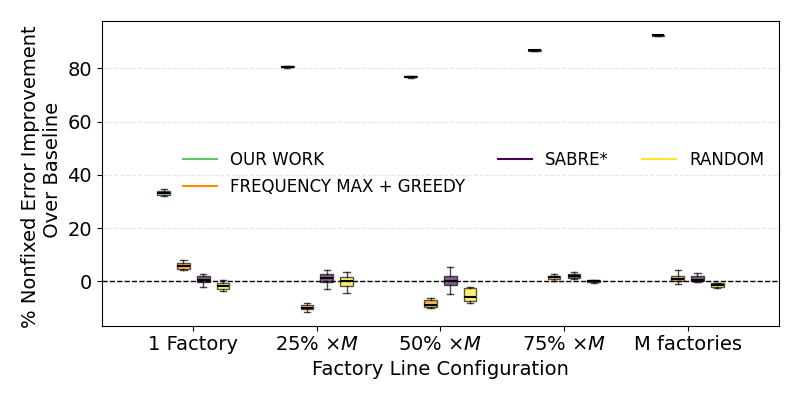}}
    \caption{\revision{adder\_433 benchmark}}\label{fig:sensitivity_c_adder}
  \end{subfigure}
  \begin{subfigure}{0.325\textwidth}
    \centering
    \revisionbox{\includegraphics[width=\linewidth]{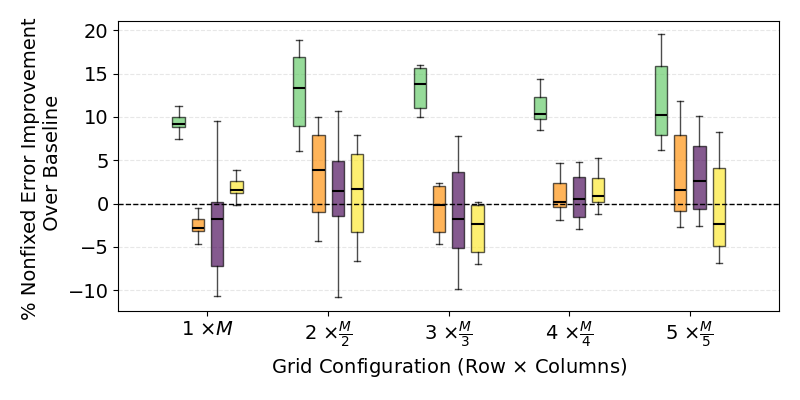}}
    \caption{\revision{knn\_341 benchmark}}\label{fig:sensitivity_c_knn}
  \end{subfigure}
  \begin{subfigure}{0.325\textwidth}
    \centering
    \revisionbox{\includegraphics[width=\linewidth]{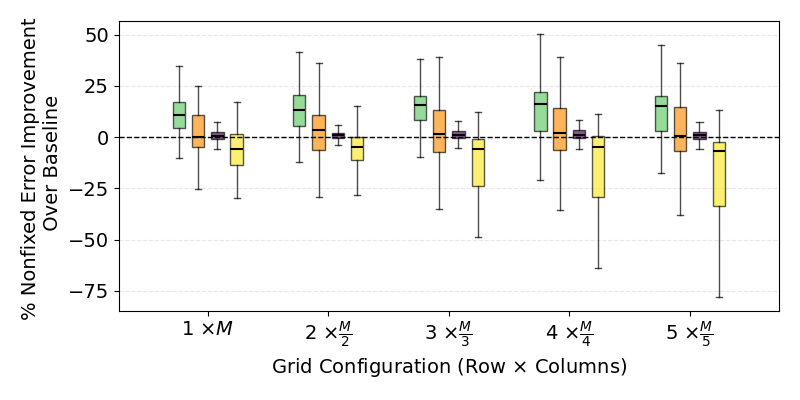}}
    \caption{\revision{Averaged across all
    benchmarks}}\label{fig:sensitivity_c_avg}
  \end{subfigure}
  \caption{{Sensitivity to grid configurations where the grid is size
      $x \times \frac{M}{x}$ where $M$ is the number of
  modules.}}\label{fig:sensitivity_c}
\end{figure*}

\revision{We show how the improvement in total failure probability
  ($P_\text{error})$ of our policy and other baselines changes depending on
  the relative contribution of errors from sources other than the
  inter-module measurements in Figure~\ref{fig:sensitivity_a}. The
  error rates used from Table~\ref{tab:error-rates} are estimates by
  IBM based on current and projected quantum device
  capabilities~\cite{tourdegross}, and correspond to a contribution of
  $1\times$, which is depicted by the \textit{base rate} line. However,
  different devices such as neutral atoms, trapped ions, improvements
  in quantum hardware, etc., as well as arbitrary extractor
  architectures, will change the relative error landscape between these
  operations. As advances are made, the $Rz$ synthesis error rates will
  reduce and therefore, FTQC architectures will shift to a regime
  corresponding to relative contributions of $< 1$. The reduction in
  $P_\text{error}$ obtained by our policy will increase as a result, thereby
making our proposal robust.}

\subsubsection{\revision{Sensitivity to Number of Logical Qubits per Module}}

\revision{We use the gross code, i.e., the $[[144,12,12]]$ code, as
  the example code throughout the paper. However, there exist many
  other candidate high-rate codes, with different $[[n,k,d]]$
  parameters. We evaluate the improvement in $P_\text{non-fixed}$ in
  Figure~\ref{fig:sensitivity_b} on varying the number of logical
  qubits per module, which will be equal to $k - 1$ for a $[[n,k,d]]$
  code. As $k$ reduces, the relative improvement in
  $P_\text{non-fixed}$ by our mapper improves since fewer qubits per module
  increases the number of modules present in the architecture, and also
  increases average number of inter-module measurements needed per
  rotation. This indicates that as we scale quantum systems and become
  capable of running circuits on a larger number of modules, the
  improvements gained by our mapping policy will become even more relevant and
  substantial, whereas the improvements obtained from the baselines
either reduce or remain the same.}

\subsubsection{\revision{Sensitivity to Factory Placement}}

\revision{We evaluate the improvements in $P_\text{non-fixed}$ of
  our mapping policy and other baselines as we allocate more
  factories within the
  architecture in Figure~\ref{fig:sensitivity_d}. We start with a
  single factory at the top as assumed by~\cite{tourdegross}, and then
  increase the factory allocation by spreading out $f\cdot M$ factories
  along the line, where $f$ is a fraction between $25\% - 100\%$ and
  $M$ is the number of modules needed by each benchmark. The
  performance of our policy improves as we increase the factory allocation,
  whereas the baselines underperform with increased factory
  availability. This is a result of not accounting for factory
placement, and only optimizing for relative placement between modules.}

\subsubsection{\revision{Sensitivity to Grid Architectures}}

\revision{We also evaluate the improvement in $P_\text{non-fixed}$ of
  our mapping policy and other baselines with respect to the grid topology as
  described in Section~\ref{sec:grid-routing}. The improvements in
  our scheme either improve or are largely unaffected by changes to the
  topology. However, the greedy and NISQ baselines start
  underperforming as we move towards wider grids since they do not
account for multiple possible placements with equivalent priorities.}

\section{Discussion and Conclusion}

The logical qubit mapping problem for the BB code is a new unexplored
problem that arises in the context of spatially efficient,
fault-tolerant quantum architectures.
Although it shares similarity with NISQ-era qubit mapping problem,
our results demonstrate that direct adaptation of NISQ mappers is insufficient.
These NISQ mappers lack the ability to adequately address the two
level-nature of the problem, are unable to model arbitrary length-$n$
interactions, and fail to distinguish between the unique clustering
and assignment subproblems.
To address these limitations, we introduce a mapping policy, which
handles the clustering problem by modeling length-$n$ logical qubit
interactions, and addresses the assignment problem using a priority
based heuristic that promotes efficient path reuse. \revision{We also
  prove the optimality of the assignment heuristic for a line
architecture and a single factory.}
Across large benchmarks, Our mapper reduces the non-fixed component
of \revision{program failure rate} by approximately $\sim 13\%$ on
average, and by $\sim 11\%$ on medium-sized benchmarks.
We also find an efficient path-finding algorithm on another specific
connection topology under consideration by researchers for fault
tolerance, the ``long grid".
Our proposal maintains strong performance across topologies achieving
an average improvement of \revision{$\sim 17\%$} in non-fixed error
relative to the baseline , demonstrating its robustness to
architectural variation.

Overall, our results indicate that we enables substantial ``free"
improvements in \revision{program failure rate ($\sim 4\%$, which
    further improves to $\sim 10.5\%$ as factory error rates will get
better)} without requiring any hardware modifications.
This alleviates part of the burden on hardware designers and provides
a promising software-level pathway toward early FTQC.

\section*{Acknowledgements}
We would like to thank Patrick Rall and Eddie Schoute for their
discussions and valuable inputs related to the tour de gross
architecture~\cite{tourdegross}. This work was funded in part by NSF
24-599: Quantum Leap Challenge Institutes (QLCI) and by the NSF STAQ
project (PHY-2325080).

\bibliographystyle{IEEEtranS}
\bibliography{refs}

\end{document}